\documentclass[preprint,12pt]{elsarticle}

\usepackage{amssymb}
\usepackage{xcolor}
\usepackage{amsmath}

\journal{Optics \& Laser Technology}

\begin{document}

\begin{frontmatter}



\title{Regimes and Transitions of the Nonlinear Temporal Talbot Effect: Underlying Mechanisms of A-Type Breathers, Soliton Crystals, and Soliton Gas}


\author[1]{Marina Zajnulina} 

\affiliation[1]{organization={Multitel Innovation Centre},
            addressline={Rue Pierre et Marie Curie 2}, 
            city={Mons},
            postcode={7000}, 
            country={Belgium}}
\ead{zajnulina@multitel.be, marina@physik.tu-berlin.de}

\begin{abstract}
A frequency comb generated from a phase-modulated continuous-wave laser is simultaneously subject to the temporal Talbot effect and modulational instability (MI) when propagating through a piece of optical fiber. The temporal Talbot effect refers to the dispersion-driven self-imaging of optical pulses and is, per se, linear in optical field amplitudes. MI is a nonlinear effect. Despite growing interest and a variety of possible applications, a concise theory of the nonlinear temporal Talbot effect that incorporates nonlinear effects is not yet available; the self-imaging of optical patterns under the influence of nonlinearity remains largely unexplored. 
Here, I derive a dispersion relation for frequency-comb spatial modes. It integrates the contributions of the linear temporal Talbot effect, self-phase modulation, and MI-driven cross-phase modulation and four-wave mixing (FWM) between frequency-comb lines, paving the way to the development of a concise theory of the nonlinear temporal Talbot effect. With the help of Soliton Radiation Beat Analysis, a technique for retrieving the soliton content of optical fields in fibers, I demonstrate that the temporal Talbot effect seeds the spatio-temporal distribution of the field by driving higher-order spatial-mode interference. Nonlinear effects modify this interference, linking Talbot effect to Fermi–Pasta–Ulam–Tsingou recurrence of optical pulses and producing input-power-dependent regime transitions between the linear Talbot effect, A-type breathers, soliton crystals, and soliton gas. I show that A-type breathers arise from regular FWM at lower input powers, whereas soliton crystals emerge from cascaded FWM at higher powers. This study advances the fields of Nonlinear Optics and Wave Theory.

\end{abstract}


\begin{highlights}
\item Unified theory of the nonlinear temporal Talbot effect is established.
\item Four input-power regimes identified and explained: linear Talbot, A-type breathers, crystals, gas.
\item Dispersion relation derived for frequency-comb spatial modes under nonlinearity.
\item Higher-order Talbot interference shown as seed of nonlinear waves and their recurrence.
\item Distinction drawn between regular-FWM breathers and cascaded-FWM soliton crystals.
\end{highlights}


\end{frontmatter}




\section{Introduction}

    A phase-modulated continuous-wave (CW) laser field has a frequency comb in its spectrum. It is one of the widely used systems in optics. For instance, it can be utilized for wave shaping \cite{Finot_2015}, LiDARs \cite{Zhang_2023}, fundamental studies of nonlinear light propagation \cite{Andral_2020} and wave collisions \cite{Friquet_2013}, and as an information carrier in frequency-multiplexed optical computing \cite{Zajnulina_2025}, \cite{Zajnulina_2025_Kerr}. 

    The frequency comb from a phase-modulated CW laser is subject to two fundamental effects in nonlinear dispersive media such as optical fibers and semiconductors. The first one is the Talbot effect, and the second one is modulational instability (MI). The Talbot effect refers to the self-imaging of a periodic optical pattern under the influence of diffraction or dispersion, both of which are linear effects in the optical field amplitudes \cite{Wen_2013}. MI, albeit depending on material dispersion, particularly anomalous dispersion, is nonlinear and associated with the dynamical growth and evolution of periodic perturbations on a CW background \cite{Dudley_2009}. Despite their simultaneous action on a frequency comb, these effects have been treated mostly separately. The combined action of these two effects is not yet fully understood, and a concise theory of the \textit{nonlinear} Talbot effect is still lacking, one that would integrate nonlinearities into the concepts of the initially linear temporal and spatial Talbot effects.
    
    Self-imaging of a spatially periodic pattern due to diffraction was first described by Henry Fox Talbot in 1836 and has since been known as the \textit{spatial} Talbot effect. It describes the reappearance of a periodic diffraction-grating image at twice the so-called Talbot period or length $L_{T}$ if illuminated by a monochromatic wave. At just $L_{T},$ the grating image reappears phase-shifted by $\pi$ \cite{Talbot_1836}. Since then, spatial Talbot effect was observed in a variety of fields: classical optics, plasmonics \cite{Dennis_2007} and X-rays \cite{Wen_2013}, Bose-Einstein condensate \cite{Zhai_2018}, lithography \cite{Solak_2011}, optical systems with gratings from acoustic waves \cite{Fisicaro_2025}, quantum mechanics \cite{Sanz_2007}, quantum-dot molecules \cite{Azizi_2021}, diatom algae \cite{Dyakov_2025}, cold atoms \cite{Kresic_2018}, ultrasonic-wave propagation in metamaterials \cite{Candelas_2019}, and many more.
    
    The \textit{temporal} Talbot effect denotes the self-imaging of a temporally periodic pattern due to the dispersion of the medium through which the pattern propagates. It was first described in 1981 \cite{Jannson_1981}, \cite{Azana_2003}. Since then, it has been used for pulse-rate multiplication \cite{Maram_2015}, \cite{Driouche_2022}, frequency combs generation \cite{Cortes_2019}, and signal processing (cf. \cite{Hyun_2024}, \cite{Yessenov_2022}) in optical fibers, generation of bright and dark pulses in semiconductors \cite{Wu_2022}, \cite{Wu_2023}, atom interferometry in Bose-Einstein condensates \cite{Wei_2024}, microwave pulse amplification \cite{Pepino_2023} and to create microwave photonic filters \cite{Maram_2019} to name a few applications.  

    The propagation of a periodic pattern in a medium becomes quickly nonlinear due to light-matter interaction. Being subject to material nonlinearity, the initially linear Talbot effect becomes \textit{nonlinear.} Studies and utilization of the nonlinear Talbot effect have attracted researchers' interest for the last two decades. Thus, nonlinear spatial Talbot effect was observed in periodically poled $\text{LiTaO}_{3}$ and $\text{LiNbO}_{3}$ crystals \cite{Wen_2011}, \cite{Liu_2014}, \cite{Yang_2018}, \cite{Wang_2021} as well as nonlinear photonic crystals \cite{Li_2021} opening a pathway for high-resolution lithography, reported as a means for excitation of nonlinear beams such as Akhmediev breathers (AB) and solitons \cite{Schiek_2021}, \cite{Cohen_2008}, as an origin of the optical-pattern formation in cold atomic clouds \cite{Labeyrie_2025}, and as a way to generate excitation gratings in the hard X-rays \cite{Goloborodko_2025}. Despite the growing interest in the nonlinear Talbot effect, spatial and temporal, it still requires further studies to be fully understood and effectively utilized in applications. Whereas some approaches have been undertaken to create a theory of the nonlinear \textit{spatial} Talbot effect (cf. \cite{Schiek_2021}, \cite{Cohen_2008}), a concise theory of the nonlinear \textit{temporal} Talbot is still missing, to the best of my knowledge.  

    MI has been extensively studied in systems governed by the Nonlinear Schr\"odinger Equation (NLS) since the 1960s \cite{Erkintalo_2011}. Those are, for instance, optical fibers, water waves, and the Bose-Einstein condensate. MI's mathematical treatment has led to a variety of NLS solutions, which have been followed by experimental observations. Thus, the fundamental MI-related solution is the Akhmediev breather (AB), a temporally periodic, spatially localized, nonlinear wave, reported in 1986 \cite{Dudley_2023}, \cite{Copie_2019}. Other solutions are spatio-temporally localized Peregrine solitons \cite{Hammani_2011}, \cite{Tikan_2017}, \cite{Ye_2020}, \cite{Chabchoub_2021} and their closely related brothers called rogue waves \cite{Akhmediev_2020}, \cite{Pan_2021}. The recognition of MI's close relationship to the Fermi–Pasta–Ulam–Tsingou (FPUT) recurrence led to finding AB recurrence under loss as a perturbation \cite{Kimmoun_2016}, construction of spatio-temporally periodic solutions under higher-order MI \cite{Erkintalo_2011}, as well as a quite recent finding and observation of a doubly periodic umbrella solution class to ABs called A-type and B-type breathers \cite{Crabb_2019}, \cite{COnforti_2020}, \cite{Vanderhaegen_2020a}, \cite{Vanderhaegen_2020b}, \cite{Vanderhaegen_2021}. 
       
   Although a frequency comb from a phase-modulated CW laser is    Although a frequency comb from a phase-modulated CW laser is simultaneously subject to both MI \cite{Dudley_2009} and temporal Talbot effect in fibers \cite{Jannson_1981}, the studies connecting these two fundamental effects to a unified framework are scarce. Thus, the nonlinear temporal Talbot effect has been brought into connection with the emergence of rogue waves from doubly periodic initial conditions \cite{Zhang_2015a}, \cite{Zhang_2015b}, \cite{Nikolic_2019}, and frequency-comb generation via cross-phase modulation (XPM) \cite{Bolger_2005}, \cite{Zhang_2022}, both MI-related effects. Further studies are needed to better understand the relationship between the temporal Talbot effect and MI and their simultaneous action, particularly concerning the spatio-temporal recurrence of optical patterns in fibers and semiconductors, to create a concise theory of the \textit{nonlinear} temporal Talbot effect. 

   In the Letter ~\cite{Zajnulina_2024}, my colleague Michael B\"ohm and I reported a strong relationship between the temporal Talbot effect and MI-related nonlinear-wave generation, studying a phase-modulated CW laser field in optical fibers. We recognized this relationship using Soliton Radiation Beat Analysis (SRBA). This numerical technique allows retrieving soliton content of nonlinear waves generated from arbitrary inputs in optical fibers \cite{Boehm_2006}, \cite{Mitschke_2017a}, \cite{Zajnulina_2015}, \cite{Zajnulina_2017}. Thus, we reported input-power-dependent transitions from a quasi-linear regime, governed primarily by linear Talbot effect at low input powers, to the regimes of the nonlinear Talbot effect of soliton crystals and separated solitons (cf. Fig.~\ref{fig:SRBA_Opt_Power}, left). These regime transitions happened at well-defined threshold input powers, giving rise to the interpretation that the NLS changes the type of its solutions at these values. 
   

    To the best of my knowledge, these were the first reported results to connect the temporal Talbot effect with the MI-related generation of soliton crystals and solitons in such detail. However, the results were derived rather heuristically from our SRBA observations. Many questions remained open, in particular, an explanation of the observed regimes and their transitions backed up by a concise theory of the nonlinear \textit{temporal} Talbot, as the latter was not available at the time. 

    The main goal of this paper is to pave the way for the development of a unified and concise theory of the \textit{nonlinear} temporal Talbot effect by integration of MI-related four-wave mixing (FWM) and cross-phase modulation (XPM) of frequency-comb lines, as well as their self-phase modulation (SPM), in the theory of the well-known \textit{linear} temporal Talbot effect. This approach enables us to better understand various input-power-dependent regimes and regime transitions of the nonlinear temporal Talbot effect, including A-type breathers, soliton crystals, and soliton gas, opening new perspectives on the nonlinear-wave formation and their recurrence under FPUT.
    
   Thus, I build upon the results of Ref.~\cite{Zajnulina_2024} and study the transitions of the temporal Talbot effect and the physical mechanisms behind them using SRBA and bridging SRBA with a derived dispersion relation for frequency-comb lines. It allows me to reveal four input-power regimes: the regime of the linear temporal Talbot effect and the nonlinear regimes of A-type breathers, soliton crystals, and separated solitons (also called soliton gas due to a weak interaction between the solitons, \cite{Suret_2024}). I show that the (higher-order) interference of the discrete spatial modes of the frequency-comb lines originates from the temporal Talbot effect and governs the spatio-temporal distribution of the optical field in all regimes, allowing for the development of an exhaustive theory of the \textit{nonlinear} temporal Talbot effect. The present study confirms the observation that frequency-comb lines can encode line-specific solitons, first reported in Ref.~\cite{Zajnulina_2024}. Additionally, I draw a difference between A-type breathers and soliton crystals. Thus, A-type breathers originate from \textit{regular} FWM at quite low input powers, whereas soliton crystals stem from \textit{cascaded} FWM at higher input powers. Exhibiting these characteristics, soliton crystals in single-pass optical fibers are no different from soliton crystals in cavities. Last but not least, I show that phase modulation depth is an important parameter that controls the type of nonlinear waves. Thus, a small modulation depth results in ABs under FPUT close to the Peregrine-soliton limit, whereas higher values of modulation depth lead to the formation of doubly periodic solutions such as A-type breathers and soliton crystals.        
    
   To conclude, for decades, MI-induced nonlinear wave generation has been considered as a separate field of studies, giving rise to different types of NLS solutions. These solutions are important for the description and classification of various nonlinear waves in optical fibers and semiconductors. However, they provide only a limited understanding of the physics behind. My present study shows that a deeper and more sophisticated insight into the dynamics of a phase-modulated CW laser field stems from the perspective of the \textit{nonlinear} temporal Talbot effect that integrates SPM and MI-related effects of FWM and XPM of frequency-comb lines in the theory of the well-known \textit{linear} temporal Talbot effect. With that, it contributes to the development of an exhaustive theory of the nonlinear temporal Talbot effect and, thus, advances the fields of Nonlinear Optics and Wave Theory.
   
   The study is structured as follows. In Sec.~\ref{sec:In_Advance_Discussion}, I present the system under study, a phase-modulated CW laser field in a single-mode optical fiber, and show that the operation happens beyond the standard AB theory. In Sec.\ref{sec:Methods}, I discuss the methodology for theoretical and numerical studies, including the SRBA. In the Results and Discussion section (Sec.~\ref{sec:Results_and_Discussion}), I present a dispersion relation for optical-frequency comb lines that includes group-velocity dispersion (GVD) and, with it, the footprint of the linear temporal Talbot effect, as well as SPM, XPM, and FWM, (Sec.~\ref{sec:SRBA_Dispersion_Relation}); then, I bridge this dispersion relation with SRBA (Sec.~\ref{sec:SRBA_Effects}); I discuss the transitions of the temporal Talbot effect from the linear to nonlinear regimes and draw the difference between the A-type breathers and soliton crystals (Sec.~\ref{sec:Transitions}); I provide a confirmation that frequency-comb lines can encode line-specific solitons (\cite{Zajnulina_2024}, Sec.~\ref{sec:Lines_Solitons}); and, finally, I discuss the impact of the modulation depth on the nonlinear-wave evolution (Sec.~\ref{sec:Modulation_Depth}). I conclude and present a summary of the achieved results in Sec.~\ref{sec:conclusion}.

   \section{In-Advance Discussion}
\label{sec:In_Advance_Discussion}
\subsection{The System Under Study: A Phase-Modulated Continuous-Wave Optical Field}
\label{sec:Setup}
\begin{figure}[htbp]
\centering
\fbox{\includegraphics[width=0.80\textwidth]{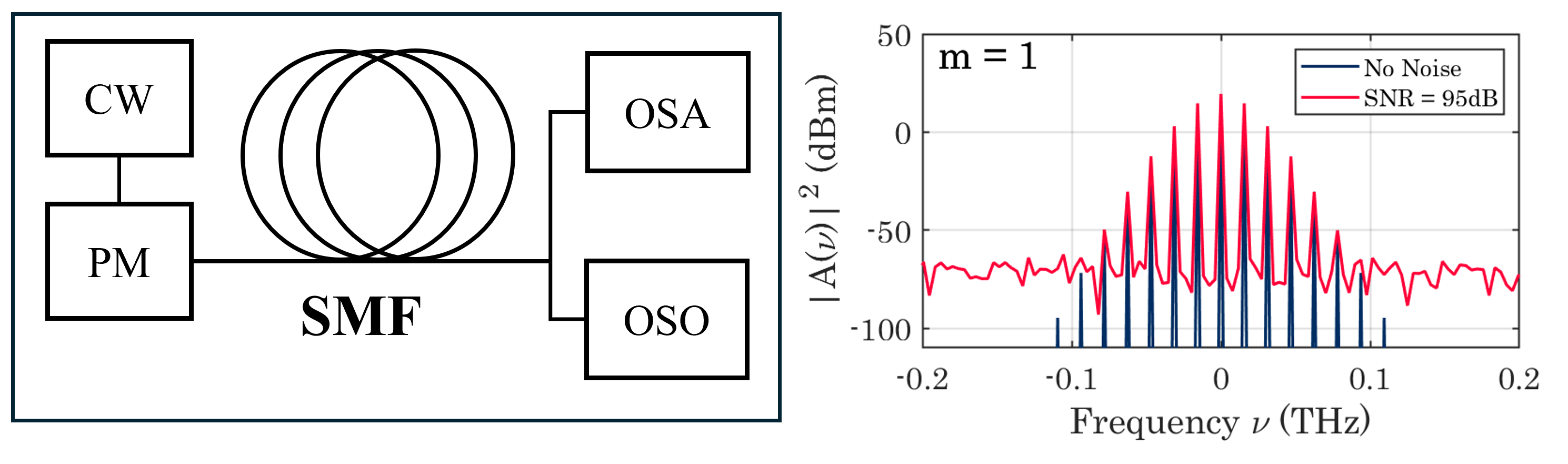}}
\caption{\textit{Left:} Schematic of the system studied here. CW: continuous-wave laser at $\lambda_{0} = 1554.6~\text{nm},$ $PM:$ phase-modulator with modulation frequency $\Omega = 15.625~\text{GHz}$ and modulation depth $m,$ SMF: single-mode fiber with GVD parameter $\beta_{2}=-23~\text{ps}^{2}/\text{km}$ and nonlinear coefficient $\gamma = 1.2~\text{(W}\cdot\text{km)}^{-1},$ OSA: optical spectrum analyzer, and OSO: optical sampling oscilloscope. \textit{Right:} Initial frequency comb with and without noise (Eq.~\ref{equ:IC}) generated by PM modulating the phase of a CW field with input power $P_0=0.15~\text{W}.$ The PM modulation depth is $m= 1$ \cite{Zajnulina_2024}.}
\label{fig:SETUP}
\end{figure}

    The optical system I study here is depicted in Fig.~\ref{fig:SETUP} (right). It consists of a CW laser with emission at a central wavelength of $\lambda_{0} = 1554.6~\text{nm}.$ The laser output is phase-modulated by the phase modulator (PM) with modulation frequency $\Omega = 15.625~\text{GHz}$ and modulation depth $m.$ The phase-modulated CW field is propagated through a single-mode fiber (SMF) and recorded by an optical sampling oscilloscope (OSA) and optical spectrum analyzer (OSA). \textcolor{black}{The nonlinear coefficient of the SMF is $\gamma = 1.2~\text{(W}\cdot\text{km)}^{-1}$ and the GVD parameter $\beta_{2}=-23~\text{ps}^{2}/\text{km}.$
    Please note that the choice of the of the fiber parameters and the value of the modulation frequency $\Omega = 15.625~\text{GHz}$ is arbitrary. In particular, $\beta_{2}$ and $\Omega$ rescale the Talbot period $L_{T}$ \cite{Zajnulina_2024}, but do not alter the existence or nature of the regimes I will discuss below.}

\begin{figure}[htbp]
\centering
\fbox{\includegraphics[width=0.90\textwidth]{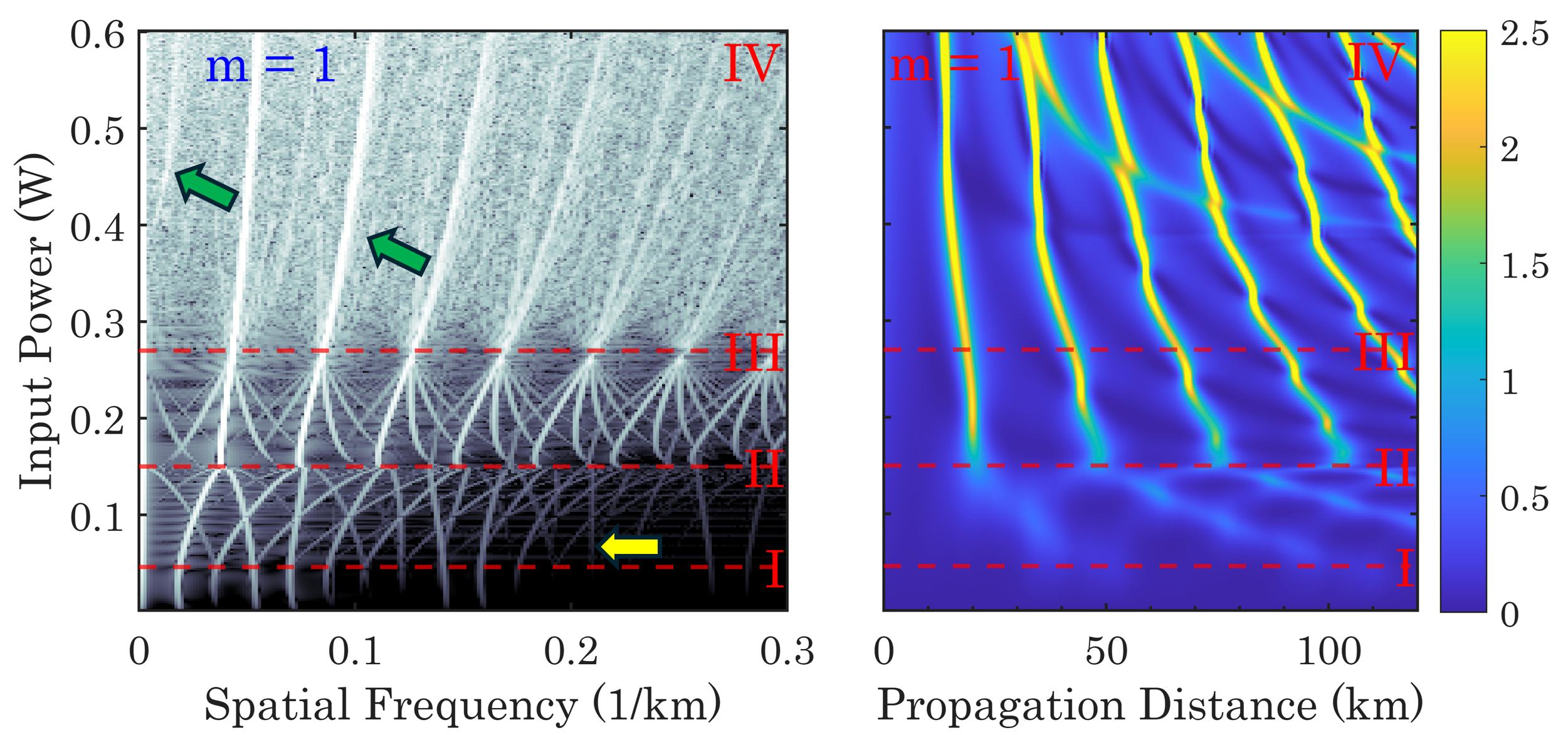}}
\caption{\textit{Left:} Soliton Radiation Beat Analysis of input-power dependent regimes in dB (cf. \cite{Zajnulina_2024}). Green arrows mark ridges of spatio-temporally separated solitons. The yellow arrow marks the fundamental Talbot frequency $Z_{T}$ (Eq.~\ref{equ:Z_Talbot}). \textit{Right:} Corresponding optical peak power $P(z) = |A(z, t= 0)|^{2}$ in W. The PM modulation depth is $\text{m}= 1.$ Dashed lines mark input-power dependent transitions of the nonlinear temporal Talbot effect.}
\label{fig:SRBA_Opt_Power}
\end{figure}

    As first discussed in Ref.~\cite{Zajnulina_2024}, there are several input-power-dependent regimes of the nonlinear temporal Talbot effect observable in this system. Thus, for the same set of fiber parameters, we reported a quasi-linear regime for input powers $P_{0}\leq0.15~\text{W},$ a regime of soliton crystals for input powers of $0.15~\text{W}<P_{0} \leq 0.27~\text{W},$ and a regime of separated solitons for $P_{0}>0.27~\text{W}$ (Fig.~\ref{fig:SRBA_Opt_Power}). An additional regime transition at $P_{0}=0.046~\text{W}$ was discovered after the publication of Ref.~\cite{Zajnulina_2024} and will be discussed in Sec~\ref{sec:Transitions}. 

    According to Ref.~\cite{Zajnulina_2024}, these regimes arise due to the temporal Talbot effect, a self-imaging effect in dispersive media such as optical fibers or semiconductors (cf. \cite{Wu_2022}, \cite{Wu_2023}). The condition for this effect is satisfied by the initial frequency comb that is provided through the phase modulation of a CW laser field \cite{Jannson_1981}, \cite{Azana_2003}, \cite{Maram_2015} (Fig.~\ref{fig:SETUP}, right). The fiber dispersion acts as a unitary phase filter, inducing the break-up of the phase-modulated CW field into a train of identical pulses \cite{Cortes_2019}. These pulses are subject to the Kerr nonlinearity of the fiber glass, which leads to the formation of the nonlinear regimes of soliton crystals and separated solitons. In Sects.~\ref{sec:SRBA_Dispersion_Relation}, \ref{sec:SRBA_Effects}, \ref{sec:Transitions}, I provide a detailed explanation of how these regimes arise, building a bridge between the linear temporal Talbot effect and the nonlinear SPM and MI-driven FWM and XPM of frequency-comb lines. It allows for the development of a concise theory of the nonlinear temporal Talbot effect.

\begin{figure}[htbp]
\centering
\fbox{\includegraphics[width=0.95\textwidth]{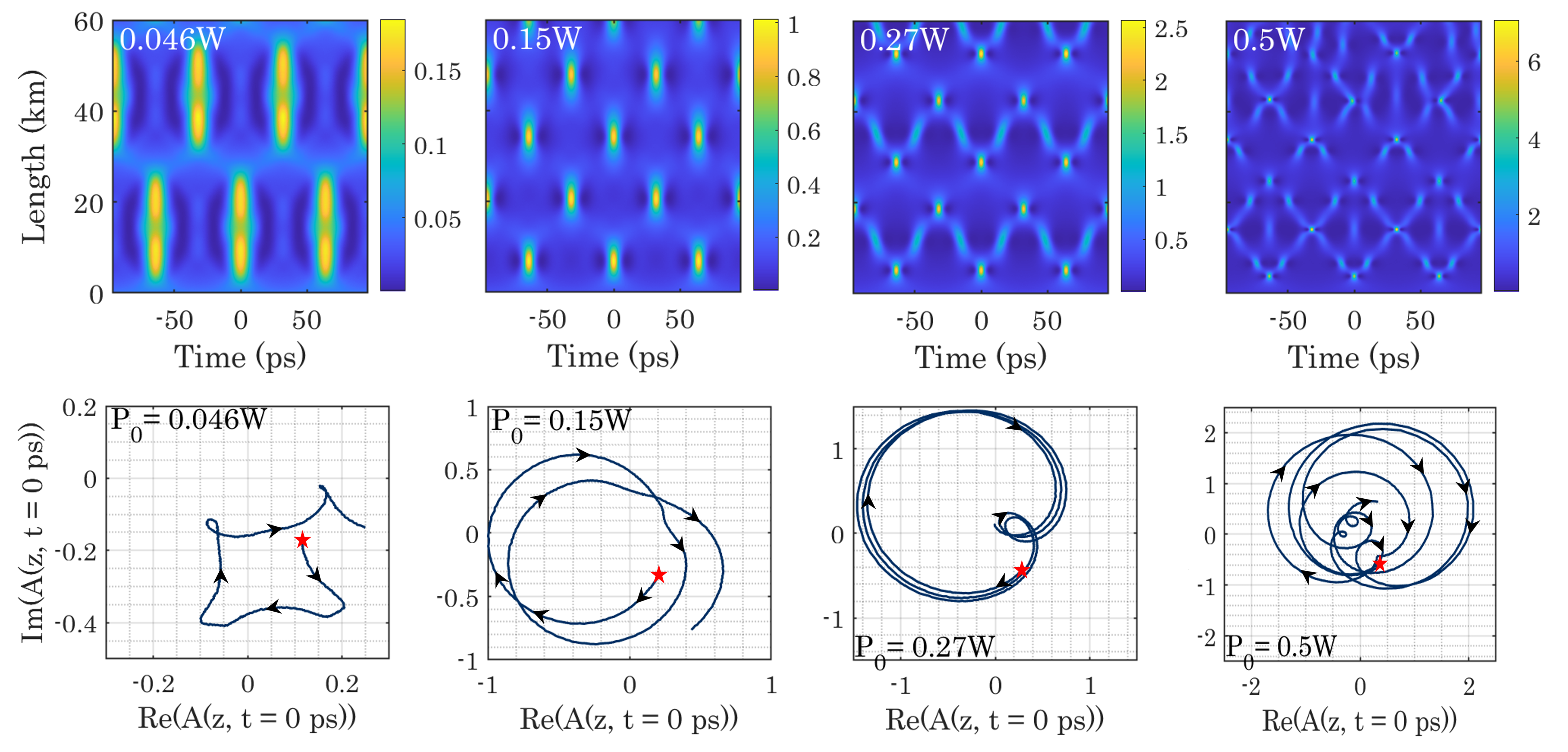}}
\caption{\textit{Top:} Optical power evolution in W along the fiber propagation distance (cf.~\cite{Zajnulina_2024}). \textit{Bottom:} Corresponding trajectories for $t = 0~\text{ps},$ i.e., the center of the chosen temporal window, with red stars denoting the starting points at $z = 0~\text{km}$ and arrows depicting the evolution direction of the trajectories. The PM modulation depth is $m= 1.$}
\label{fig:Trajectories}
\end{figure}

\subsection{Operation Beyond the Established Akhmediev Breather Theory}
\label{sec:Why_Not_AB}
\begin{figure}[htbp]
\centering
\fbox{\includegraphics[width=0.98\textwidth]{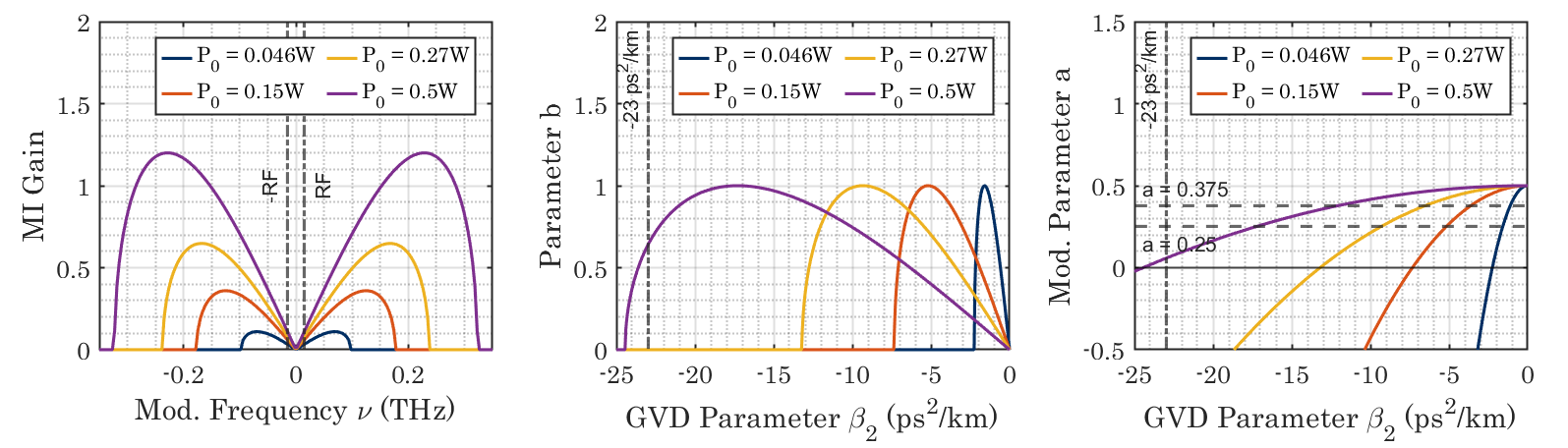}}
\caption{\textit{Left:} Modulational instability (MI) gain in $1/\text{km}$ for input powers $P_{0} = 0.046~\text{W},$ $0.15~\text{W},$ $0.27~\text{W},$ and $0.5~\text{W}.$ \textit{Middle:} Akhmediev breather parameter $b$ as a function of group-velocity dispersion (GVD) parameter $\beta_{2}.$ \textit{Right:} Akhmediev breather parameter $a$ as a function of group-velocity dispersion (GVD) parameter $\beta_{2}$ for the same values of input power.}
\label{fig:AB_Parameters}
\end{figure}

    As discussed above, a frequency comb of a phase-modulated CW field turns into a train of optical pulses due to the temporal Talbot effect (\cite{Jannson_1981}, \cite{Azana_2003}, \cite{Maram_2015}, \cite{Cortes_2019}). This pulse train is subject to MI due to the Kerr nonlinearity of the fiber glass. MI denotes a growth of initially weak modulation at the expense of the CW background \cite{Dudley_2009}. The MI gain $g(\Omega)$ in $1/\text{km}$ is given by \cite{Agrawal_2019}:
\begin{equation}
    g({\Omega}) = |\beta_{2}\Omega|\sqrt{ \frac{4\gamma P_{0}}{|\beta_{2}|} - \Omega^{2}}
\end{equation}
with $\Omega$ being the modulation frequency.

    An AB is a temporally periodic and spatially localized wave that arises due to MI. It is a well-known analytical solution of the NLS (Eq.~\ref{equ:NLS}) \cite{Dudley_2023}, \cite{Copie_2019}. Under a small perturbation, for instance, by a fiber loss, it enters the regime of FPUT recurrence over the propagation distance \cite{Kimmoun_2016}. The optical power distribution over time and space has a strong similarity to the optical power evolution we observed in the regime of what we call soliton crystals \cite{Zajnulina_2024}. This gives rise to the interpretation that our soliton crystals are in fact ABs under FPUT recurrence. However, our results were produced without taking the optical loss into account. Also, the low-level additive white noise in the initial condition (Eq.~\ref{equ:IC}) was not a reason for the FPUT recurrence, which was checked separately. No other small perturbations were included. Therefore, they cannot be considered as a reason for the spatio-temporal periodicity (or recurrence) of optical pulses in the regimes of soliton crystals and separated solitons reported in Ref.~\cite{Zajnulina_2024}.

    Fig.~\ref{fig:AB_Parameters} shows the MI gain $g(\Omega)$ and the corresponding AB parameters $a$ and $b$ for input powers studied in Ref.~\cite{Zajnulina_2024} and 
    in more detail below. The AB parameters are fundamental to the AB solution of the NLS and are defined as $2a=(1-(\Omega/\Omega_{c})^{2})$ and $b=\sqrt{8a(1-2a)}$ with $\Omega_{c}= 4\gamma P_{0}/|\beta_{2}|$ such that the maximum gain happens for $b=1$ and $a=1/4$ \cite{Dudley_2009}. For $\beta_{2}= -23~\text{ps}^{2}/\text{km}$ used here, the AB parameters are $\leq0$ and, thus, not satisfactory for an AB solution for the most significant input power values of $P_{0}=0.046~\text{W},$ $0.15~\text{W},$ and $0.27~\text{W}$ (cf. Fig.~\ref{fig:Trajectories}, top). Also, the higher-order MI does not come into question as a driving mechanism for our observations, as it requires a value of $a\geq 0.375$ \cite{Erkintalo_2011} which is achievable only for low absolute values of GVD. Apart from that, the PM modulation frequency RF of $\Omega = 15.625~\text{GHz}$ generates only a little MI gain far beyond its input-power dependent maxima. From these observations, I conclude that the results reported in Ref.~\cite{Zajnulina_2024} and studied here in more detail do not originate from ABs under FPUT recurrence despite their visual similarity (\cite{Andral_2020}). Rather, they are the result of the \textit{nonlinear} temporal Talbot effect that incorporates SPM and MI-driven FWM and XPM of frequency-comb lines as discussed in Sec.~\ref{sec:Results_and_Discussion}.  

    A more generalized class of NLS solutions constitutes the spatio-temporally periodic A-type and B-type breathers (\cite{COnforti_2020}, \cite{Vanderhaegen_2020a}, \cite{Vanderhaegen_2020b}). In Ref.~\cite{Vanderhaegen_2021}, the authors discussed that A-type breathers exist beyond the standard MI bandwidth of modulation frequencies, i.e., for higher modulation frequencies than the cut-off frequency of $g(\Omega)$ (Fig.~\ref{fig:AB_Parameters}, left), which gives rise to "extraordinary" MI. Below, I will discuss a regime of the nonlinear temporal Talbot effect for input powers of $0.046~\text{W} < P_{0}\leq 0.15~\text{W}$ that hosts doubly-periodic waves with properties of A-type breathers (Sec.~\ref{sec:Transitions}). Apparently, the PM modulation frequency does not need to exceed the cut-off frequency of $g(\Omega) $. It is also possible to generate A-type breathers within the MI bandwidth for a small value of $g(\Omega).$

    Soliton crystals, spatio-temporally periodic soliton compounds for $0.15~\text{W}<P_{0}\leq 0.27~\text{W},$ constitute a separate class of NLS solutions. The difference to A-type breathers lies in the underlying physics. As discussed in Secs.~\ref{sec:SRBA_Dispersion_Relation}, \ref{sec:SRBA_Effects}, \ref{sec:Modulation_Depth}, the A-type breathers are generated by MI-driven $\textit{regular}$ FWM at quite low input powers and small PM modulation depths; soliton crystals originate from \textit{cascaded} FWM at higher input powers and strong PM modulation depths. For higher input powers, $P>0.27~\text{W},$ soliton crystals dissolve into spatio-temporally separated solitons, i.e., soliton gas (cf. \cite{Suret_2024}).   

    I will show that higher-order interference between the spatial modes of the frequency-comb lines driven by the temporal Talbot effect is the underlying mechanism for any type of the discussed NLS solutions (Sec.~\ref{sec:Transitions}). Therefore, I conclude that an exhaustive consideration of nonlinear-wave evolution in optical fibers should be conducted from the perspective of the nonlinear temporal Talbot effect, rather than considering the linear temporal Talbot effect and SPM, as well as MI-driven effects of FWM and XPM separately, as usually done. Now, let us dive into more. 

\section{Methods}
\label{sec:Methods}
\subsection{Modelling of Light Propagation in Fibers}
\label{sec:GLNS}
The modelling of the nonlinear light propagation in the SMF (Fig.~\ref{fig:SETUP}) is done via the Nonlinear Schr\"odinger Equation (NLS) for the optical field amplitude $A(z,t)$ in the slowly varying envelope approximation in the co-moving frame \cite{Agrawal_2019}:
\begin{equation}
   \frac{\partial A}{\partial z} = -i\frac{\beta_2}{2}\frac{\partial^{2}A}{\partial t^{2}} + i\gamma |A|^{2}A.
   \label{equ:NLS}
\end{equation}
    with $\beta_2 = -23~\text{ps}^{2}/\text{km}$ being the GVD parameter and $\gamma = 1.2~\text{W}^{-1}\cdot\text{km}^{-1}$ the nonlinear coefficient at CW laser wavelength of $\lambda_{0} = 1554.6~\text{nm}.$ To concentrate on the understanding of the relationship between the temporal Talbot effect and Kerr-induced nonlinear effects, I neglect higher-order dispersion, the shock of utlrashort pulses, Raman effect, and optical losses usually present in standard SMFs \cite{Agrawal_2019}.

    The corresponding initial condition reads as:
\begin{equation}
    A(z= 0, t) = \sqrt{P_{0}}e^{i{\left(\omega_{0}t + m\cos{(2\pi\Omega t)} \right)}} + \sqrt{n_{0/\text{rand}}(t)}e^{i\phi_{\text{rand}}(t)}
    \label{equ:IC}
\end{equation}
    with $P_{0}$ being the input power provided by the CW laser. The second term represents white noise with $n_{0/\text{rand}}$ random noise amplitude proportional to $2P_{0}e^{-10}$ and random phase in the range of $\phi_{\text{rand}}\in[0, 2\pi].$ It is added to make the initial condition more realistic as compared to a fully noiseless input. The level corresponds to the spectral signal-to-noise ratio of $\text{SNR} = 95~\text{dB}$ (Fig.~\ref{fig:SETUP}, right). Please note that the additive white noise is omitted in the theoretical considerations in the following sections (Secs.~\ref{sec:SRBA_Dispersion_Relation}, \ref{sec:Transitions}), but is used to produce numerical simulation results. Its quite low (but still realistic) value does not impact the nonlinear dynamics studied in the following, which was checked separately.  

    The first term in Eq.~\ref{equ:IC} can be rewritten using the Jacobi–Anger expansion as
\begin{equation}
A(z=0, t) = \sqrt{P_{0}}\sum_{k = -\infty}^{+\infty} i^k J_{k}(m)e^{i{\left(\omega_{0} + 2\pi k \Omega \right)}t} + \text{noise}
\label{equ_IC_JA}
\end{equation}
    with $J_{k}(m)$ being modulation-depth-dependent Bessel functions of the first kind. It is a frequency comb centered at central frequency $\omega_{0}$ with a line spacing of $2\pi\Omega.$ With that, it satisfies the condition of the linear temporal Talbot effect \cite{Jannson_1981} (Fig.~\ref{fig:SETUP}, right). 

    The numerical integration of Eqs.~\ref{equ:NLS} and \ref{equ:IC} was performed using the Fourth-Order Runge–Kutta in the Interaction Picture Method (\cite{Hult_2007}), including local error estimation to achieve more accurate results, particularly in the nonlinear regimes. 

\subsection{Soliton Radiation Beat Analysis}
\label{sec:SRBA_General_Theory}

    Soliton Radiation Beat Analysis (SRBA) is a numerical technique that allows for the retrieval and quantification of the soliton content of optical fields evolving in optical fibers under the impact of the Kerr nonlinearity. The strength of this technique lies in its applicability to arbitrary inputs \cite{Boehm_2006}, \cite{Mitschke_2017a}. 

    SRBA analyzes the spatial frequencies of the oscillations along the fiber that arise due to the beating, i.e., interference, between solitons or between solitons and dispersive-wave background and, thus, allows revealing the formation and content of such nonlinear waves as solitons, breathers, and solitons crystals \cite{Zajnulina_2024}, \cite{Boehm_2006}, \cite{Zajnulina_2015}, \cite{Zajnulina_2017}, \cite{Mitschke_2017a}.

    Conceptually, SRBA exploits the knowledge of phase accumulation during nonlinear light propagation in the fiber. Thus, solitonic waves (breathers, crystals, and separated solitons) accumulate nonlinear phase proportional to optical power, $\phi_{\text{nl}}(z) \sim \gamma P z,$ whereas linear (dispersive) waves accumulate phase according to GVD, $\phi_{\text{lin}}(z) \sim \beta_2 \Omega^2 z.$ The different phases cause interference effects that manifest themselves as spatial beat patterns in the propagated optical power. SRBA extracts and analyzes these beat frequencies to infer the underlying wave structure.

    \textcolor{black}{Performing \textit{standard} SRBA involves five steps with theoretical background and a detailed implementation guide described in Refs.~\cite{Boehm_2006},\cite{Mitschke_2017a}, \cite{Zajnulina_2015}. Here, I provide a summary needed to better undertand further stages of this study.}

\begin{enumerate}[i)]
  \item \textit{Numerical Simulation:}
    Propagate an initial optical field \( A(z=0, t) \) through an NLS (Eq.~\ref{equ:NLS}) with realistic parameters \(\beta_2\) and \(\gamma\). As the resolution of SRBA images scales with the fiber length, sufficiently long propagation distances \( z \), typically several hundreds of kilometers, are required to obtain images with well-resolved spatial frequency structure. \textcolor{black}{Throughout this study, I use a fiber length of $L=500~\text{km}$ for SRBA images. This value is a trade-off between the fiber length required for high-resolution SRBA images and computational time.}

  \item \textit{Extraction of a Temporal Slice:}
  Choose a fixed-time slice of $A(z,t)$ (for simplicity, \( t = 0 \) at the center of the temporal window) and record the optical power along the propagation distance:
  \[
  P(z) = |A(z, t=0)|^2
  \]
  This function captures the longitudinal intensity variation caused by interference between solitons and dispersive radiation or due to the formation of breathers or soliton crystals.

  \item \textit{Apodization:}
  Apply a smooth apodization function \( w(z) \) to reduce the discontinuity at the boundaries and emphasize the region of interest:
  \[
  P_{\text{apo}}(z) = w(z) \cdot P(z)
  \]
  In the context of this study, I use a Gaussian as an apodization function $w(z)$.

  \item \textit{Fourier Transform:}
  Perform a fast Fourier transform (FFT) on the apodized power:
  \[
  \tilde{P}(Z) = \text{FFT}[P_{\text{apo}}(z)]
  \]
  Here, \( Z \) has units of inverse propagation distance, i.e., \( \text{km}^{-1} \), and represents beat frequencies due to phase mismatches between field components. Please note that not only beating frequencies are picked up by the FFT, but any spatially oscillatory behavior of the optical power. Thus, the temporal Talbot effect will cause a regular oscillation over the propagation distance and, therefore, generate a spatial-frequency marker called fundamental Talbot frequency $Z_{T}$ and discussed below (Eq.~\ref{equ:Z_Talbot}).  

  \item \textit{Construction of the SRBA Image and Interpretation:}
  Repeat steps i)--iv) for a range of input powers $P_{0}$ and plot the spectral amplitude \( |\tilde{P}(Z)|^2 \) in dB as a function of spatial frequencies \( Z \) and $P_{0}$, producing a two-dimensional SRBA plot (Fig.~\ref{fig:SRBA_Opt_Power}, right).

  To interpret SRBA plots, one needs to consider the shapes of the SRBA ridges. 
  \begin{itemize}
    \item \textit{Linear waves:} Input-power independent vertical ridges with constant \(Z\) correspond to the beating of linear, dispersive-wave components of the optical field (\cite{Boehm_2006}).
    \item \textit{Separated solitons:} Power-dependent rather parabolic ridges \(Z(P_0)\) correspond to emergence and evolution of solitons (\cite{Boehm_2006}, \cite{Mitschke_2017a}). 
    \item \textit{Compounds of solitonic waves:} Spatio-temporally periodic compounds such as breathers or soliton crystals manifest themselves in bundles of multiple ridges arising at different spatial frequencies with an input-power dependent intra-bundle relationship. Thus, this intra-bundle relationship constitutes bundles of pitch-fork (or lawn-rake) SRBA ridges in the case of breathers, whereas soliton crystals exhibit rather fans of spatial frequencies. This interpretation of SRBA compound-related ridges was discussed heuristically by my colleagues and me in Refs.~\cite{Zajnulina_2015}, \cite{Zajnulina_2017}, \cite{Zajnulina_2024}. This paper presents a theoretical analysis of SRBA breather and crystal bundles, confirming previous results obtained heuristically (Secs.~\ref{sec:SRBA_Dispersion_Relation}, \ref{sec:SRBA_Effects}).
  \end{itemize}
\end{enumerate}

    Here, I introduce a \textit{time-resolved} SRBA that considers spatial frequencies over time for fixed values of input power $P_{0}$ (Fig.~\ref{fig:SRBA_over_Time}). It allows for revealing the temporal localization of the SRBA ridges and thus contributes to a better understanding of physical effects such as self-phase modulation (SPM), cross-phase modulation (XPM), and four-wave mixing (FWM) of frequency comb lines (Sec.~\ref{sec:SRBA_Effects}). In this SRBA representation, the rides corresponding to solitons and soliton beating are \textit{horizontal} to the x-axis of time (cf~\cite{Hartwig_2010}, \cite{Zajnulina_2024}).  

    To conclude, SRBA provides a way to quantify solitons and solitonic-wave types (separated solitons vs. compounds such as breathers and crystals). With that, it helps detect transitions from one type of solitonic wave to another (\cite{Zajnulina_2024}). With the power of modern computers, it is computationally inexpensive and highly adaptable, and, most importantly, does not rely on the integrability of the wave equation, allowing for studies of real-world fiber systems with arbitrary inputs. Although the optical power $P(z) = |A(z, t= 0)|^{2}$ in Fig.~\ref{fig:SRBA_Opt_Power} (right) indicates the existance of different regimes, it does not provide any information about the type of solitonic waves involved as compared to its spatial Fourier spectrum, i.e., the neighboring SRBA image. Therefore, in the following, I will focus mainly on the SRBA plots rather than on the optical-power representation of light evolution in the SMF (Fig.~\ref{fig:SETUP}, left).

\section{Results and Discussion}
\label{sec:Results_and_Discussion}
\subsection{Dispersion Relation of the Nonlinear Schr\"{o}dinger Equation}
\label{sec:SRBA_Dispersion_Relation}

    To better understand the meaning of various ridges in Fig.~\ref{fig:SRBA_Opt_Power}, left, and the physical effects behind them, let us here derive a dispersion relation for the NLS (Eq.~\ref{equ:NLS}) with a phase-modulated CW field as initial condition (Eq.~\ref{equ:IC}). For this, I use the ansatz that expands the optical field as a Jacobi–Anger series (cf. Eq.~\ref{equ_IC_JA}) and allows us to study the phase accumulation along the propagation variable $z:$
\begin{equation}
    A(z, t) = \sqrt{P_{0}}e^{i\zeta\cdot z}\sum_{k = -\infty}^{+\infty} i^k J_{k}(m)e^{i{\left(\omega_{0} +  2\pi k\Omega \right)}t}
\label{equ:NLS_ansatz}
\end{equation}
    and plug it into the NLS (Eq.~\ref{equ:NLS}). Please, keep in mind that this ansatz treats the initial comb from a phase-modulated CW laser field as a sum of monochromatic waves with frequencies $\omega_{k}=\omega_{0} + 2\pi k\Omega,$ $k\in\mathbb{Z}.$

    With Eq.~\ref{equ:NLS_ansatz}, the left-hand side of the NLS (Eq.~\ref{equ:NLS}) delivers:
\begin{equation}
    \frac{\partial A}{\partial z} = i \zeta \sqrt{P_0} \, e^{i \zeta z} \sum_{k=-\infty}^{\infty} i^k J_k(m) e^{i(\omega_0 + 2\pi k \Omega)t} = i \zeta A(z, t).
    \label{equ:LHS}
\end{equation}
The dispersive term on the right-hand side gives:
\begin{equation}
    - i \frac{\beta_2}{2} \frac{\partial^2 A}{\partial t^2}
= i \frac{\beta_2}{2} \sqrt{P_0} \, e^{i \zeta z} \sum_{k=-\infty}^{\infty} i^k J_k(m) \omega_k^2 e^{i \omega_k t}.
    \label{equ:RHS_Dispersion}
\end{equation}

    The nonlinear term on the right-hand side of Eq.~\ref{equ:NLS} results in:
\begin{equation}
i\gamma|A|^2 A = i\gamma P_0^{3/2} e^{i\zeta z} \sum_{i,l,j=-\infty}^{+\infty} i^{i-l+j} J_i(m)J_l(m)J_j(m) e^{i\omega_{i-l+j}t}
 \label{equ:RHS_Nonlinearity}
\end{equation}

The dispersion relation is obtained by combining coherent terms (Eqs.~\ref{equ:LHS}, \ref{equ:RHS_Dispersion}, \ref{equ:RHS_Nonlinearity}):
\begin{equation}
\zeta_k(\omega_k) = 
\underbrace{\frac{\beta_2}{2}\omega_k^2}_{\text{GVD}}
+ \gamma P_0[
\underbrace{J_k^2}_{\text{SPM}}
+ \underbrace{2\sum_{n \neq k} J_n^2}_{\text{XPM}}
+ \underbrace{\frac{1}{J_k}\sum_{\substack{i,l,j \\ i-l+j=k \\ i \neq l}} J_i J_l J_j}_{\text{FWM}}
]
\label{equ_DispersionRalation_Full}
\end{equation}

    Here, follows the first important result: due to a sum of optical frequencies $\omega_{k}$ on the right hand-side of Eq.~\ref{equ_DispersionRalation_Full}, $\zeta$ becomes discrete splitting in the spatial modes $\zeta_{k}(\omega_{k})$ of the optical field $A(z,t).$ Also, I separated out the XPM terms from the nonlinear mixing sums in Eq.~\ref{equ:RHS_Nonlinearity}. Those are not XPM terms in the standard understanding of a nonlinear process when two signals propagate simultaneously through an optical fiber and mutually influence each other's phase. Instead, their terms describe how frequency-comb lines cross-phase modulate each other with their intensities. These XPM terms arise due to treating the initial condition (Eq.~\ref{equ:IC}) and the ansatz (Eq.~\ref{equ:NLS_ansatz}) as a sum of monochromatic waves $\omega_{k}$ and will be useful for explanation of the SRBA observations in Secs.~\ref{sec:SRBA_Effects} and \ref{sec:Transitions}.   

The general solution of the NLS reads then as:
\begin{equation}
    A(z, t) = \sqrt{P_{0}}\sum_{k = -G}^{+k} i^k J_{k}(m)e^{i\zeta_k\cdot z}e^{i\omega_{k}t}, \quad \omega_{k}=\omega_{0} +  2\pi k\Omega.
\label{equ:NLS_solution_general}
\end{equation}

    To conclude, the spatial modes Eq.~\ref{equ_DispersionRalation_Full} of the optical field $A(z,t)$ are i) discrete due to the discreteness of the optical spectrum, ii) are subject to line-specific linear dispersion, iii) self-modulate themselves due to their intensity, iv) are subject to XPM by the intensities of the other comb lines, and v) depend on FWM between the comb lines.    

\subsection{Soliton Radiation Beat Analysis Explained by the Dispersion Relation of the Spatial Modes of the Optical Field}
\label{sec:SRBA_Effects}

Eq.~\ref{equ_DispersionRalation_Full} shows the physical effects behind the spatial-mode evolution of the optical field $A(z,t).$ Now, I use this information to gain insight into the ridges of SRBA (Fig.~\ref{fig:SRBA_Opt_Power}). The following relations can be established.

\textbf{Spatial Modes of $A(z,t)$:} $\zeta_{k}(\omega_k)$ are field-own (nonlinear) complex propagation constants (spatial modes) influenced by GVD, SPM, XPM, and FWM. They are not directly observable. Rather, they govern the internal structure of the optical field $A(z,t).$
  
\textbf{SRBA Ridges:} $Z_{nm}$ are peaks in the spatial Fourier transform of the optical power $P(z)=|A(z,t= 0)|^2$ that arise due to the beating or, in other words, interference, between field's spatial modes (cf. \cite{Boehm_2006}, \cite{Mitschke_2017a}): 
  \begin{equation}
      P(z)=|A(z,t= 0)|^2 = P_{0}\sum_{n,m}i^{n-m}J_{n}(m)J_{m}(m)e^{i(\zeta_{n}(\omega_{n}) - \zeta_{m}(\omega_{m}))z}
      \label{equ:Power_Beating}
  \end{equation}
  with oscillating terms at $z$ for
  \begin{equation}
      Z_{nm}=|\zeta_{n}(\omega_{n})-\zeta_{m}(\omega_{m})|
      \label{equ:Z_nm}
  \end{equation}

with spatial modes $\zeta_{k}(\omega_{k})$ as given by the dispersion relation Eq.~\ref{equ_DispersionRalation_Full}. 

The quantities $Z_{nm}$ can be observed as fans, pitchfork-shaped, or parabolic ridges in the SRBA plots. Each field mode $\zeta_{0}(\omega_{0}),$ $\zeta_{\pm 1}(\omega_{\pm 1})$, $\zeta_{\pm 2}(\omega_{\pm2})$ etc., gives rise to a separate ridge in the SRBA plot. The ridges move non-uniformly with $P_{0}$, depending on their Bessel function amplitudes and interaction terms, which results in an unequal spacing between them.

    \textbf{Group-Velocity Dispersion (GVD):} $\zeta_{k}(\omega_{k})\propto\frac{\beta_2}{2} \omega_k^2$ is the intrinsic spatial frequency in the absence of the nonlinearity. It is determined by the \textit{linear} temporal Talbot effect (\cite{Jannson_1981}, \cite{Azana_2003}, \cite{Maram_2015}, \cite{Cortes_2019}). In the SRBA plot, it seeds the starting positions of the ridges for $P_{0}\rightarrow 0~\text{W} $.
  
  \textbf{Self-Phase Modulation (SPM):} $\zeta_{k}(\omega_{k})\propto \gamma P_0 J_k^2$ describes the nonlinear phase shift that the spatial mode $\zeta_{k}(\omega_{k})$
  experiences due to its intensity (\cite{Agrawal_2019}). In the SRBA plot, it causes the ridge associated with $\omega_{k}$ to move to higher spatial frequencies $Z$ with increasing input power $P_{0}.$ In the temporal domain, it implies that the recurrence of pulse trains happens at shorter propagation distances as seen in Fig.~\ref{fig:Trajectories}, top (cf. \cite{Dudley_2009}, \cite{Zajnulina_2024}). 
  
  \textbf{Cross-Phase Modulation (XPM):} $\zeta_{k}(\omega_{k})\propto2\gamma P_0 \sum_{n \neq k} J_n^2$ describes the nonlinear phase shift of the spatial mode $\zeta_{k}(\omega_{k})$ due to the influence of other frequency-comb components at $\omega_{n\neq k}$ (\cite{Agrawal_2019}, \cite{Bolger_2005}). 
  As the comb line intensities are not homogeneous (cf. Fig.~\ref{fig:SETUP}, right), XPM leads to non-uniform shifts in the spatial frequencies of different SRBA ridges. This, in turn, introduces a curvature of SRBA ridges and changes their spacing relative to each other as the input power $P_{0}$ increases. Apart from that, XPM contributes to the broadening of the initial comb by changing the nonlinear phase shift that determines phase matching for FWM (cf. \cite{Zhang_2022}, \cite{Bolger_2005}, cf. \cite{Yan_2024}). 

  \textbf{Four-Wave Mixing FWM:} $\zeta_{k}(\omega_{k})\propto \frac{\gamma P_{0}}{J_k}\sum_{\substack{i,l,j \\ i-l+j=k, i \neq l}} J_i J_l J_j$ describes the generation of new frequency-comb components due to FWM. This interaction leads to the modulation of the phase of each participating spatial-mode component. Over distance, it pulls the components into a collective behavior, locking their phases such that $\zeta(\omega_{k})\approx \zeta(\omega_{k\pm1 })\approx \zeta(\omega_{k\pm2})\approx\dots.$ This phase locking gives rise to coherent structures such as breathers or soliton crystals (cf. \cite{Dudley_2009}, \cite{Zajnulina_2024}). In the SRBA plot, the mode phase locking causes the SRBA ridges to merge, cluster, or evolve together, forming pitchforks and fan-like structures. FWM contributes to the spectral broadening while temporarily compressing the pulses (\cite{Dudley_2009}, \cite{Agrawal_2019}). 

  In a \textit{regular} FWM process, a nonlinear interaction between three spectral components $\omega_{n},$ $\omega_{m}$ and $\omega_{\ell}$ generates a fourth component, $\omega_{k}.$ This process yields a limited number of new frequency components, all depending on the number and phase matching of input frequencies (\cite{Agrawal_2019}). 
  A \textit{cascaded} four-wave mixing is a sequential process where the FWM products generated by an initial FWM interaction act as inputs for further FWM processes. It yields many additional frequency components, allowing for the generation of broadband frequency combs through a cascade of FWM processes (cf. \cite{McKinstrie_2006}, \cite{Li_2012}, \cite{Junaid_2024}). \textcolor{black}{Please note that there is no strict definition of \textit{regular} and \textit{cascaded} FWM, and no clear difference between them is available in the literature. In its core, it is the same wave-mixing process governed by the Kerr nonlinearity for both types of FWM. The difference is rather made based on the role of generated waves: in \textit{cascaded} FWM, newly generated comb lines serve as a source of further FWM products, whereas in \textit{regular} FWM, they do not.}
  
  For a satisfied phase-matching condition, the input power $P_{0}$ decides over the type of FWM. Thus, higher input power enables the transition from regular to cascaded four-wave mixing by amplifying new frequencies to levels sufficient for further nonlinear interactions, favoring the cascaded process \cite{Agha_2009}, \cite{Weigland_2015}. As shown in Sec.~\ref{sec:Modulation_Depth}, the PM modulation depth $m$ is another important parameter that is decisive for the type of FWM. The difference between the type of FWM becomes important in the differentiation between the A-type breathers and soliton crystals in Sec.~\ref{sec:Transitions}.
  
  \textbf{Spatio-Temporally Separated Solitons/Soliton Gas:} With increasing input power $P_{0},$ the common action of SPM and XPM deteriorates the phase-matching condition needed for FWM,  breaking down the collective phase-locked state. With that, the system transforms from a crystalline structure to spatio-temporally separated solitons \cite{Zajnulina_2015}, \cite{Zajnulina_2024}, also referred to as soliton gas due to a weak interaction of the solitonic pulses \cite{Suret_2024}. In SRBA, this is seen as ridge bundles dissolving into individual branches with quite parabolic shapes that are footprints of solitons \cite{Boehm_2006}. 

    To conclude, the dispersion relation of spatial modes ($\zeta_{k}(\omega_{k})$ allows us to provide a physical interpretation of SRBA ridge patterns. It explains how GVD, SPM, XPM, and FWM modify the spatial modes, influencing SRBA ridge positions, curvature, and merging in the SRBA plots. Additionally, it highlights how increasing input power drives transitions from weakly interacting comb lines to phase-locked states, and ultimately to spatio-temporally localized solitons.

\subsection{Transitions of the Temporal Talbot Effect: Linear Talbot Effect, A-Type Breathers, Soliton Crystals, and Soliton Gas}
\label{sec:Transitions}
\begin{figure}[htbp]
\centering
\fbox{\includegraphics[width=0.98\textwidth]{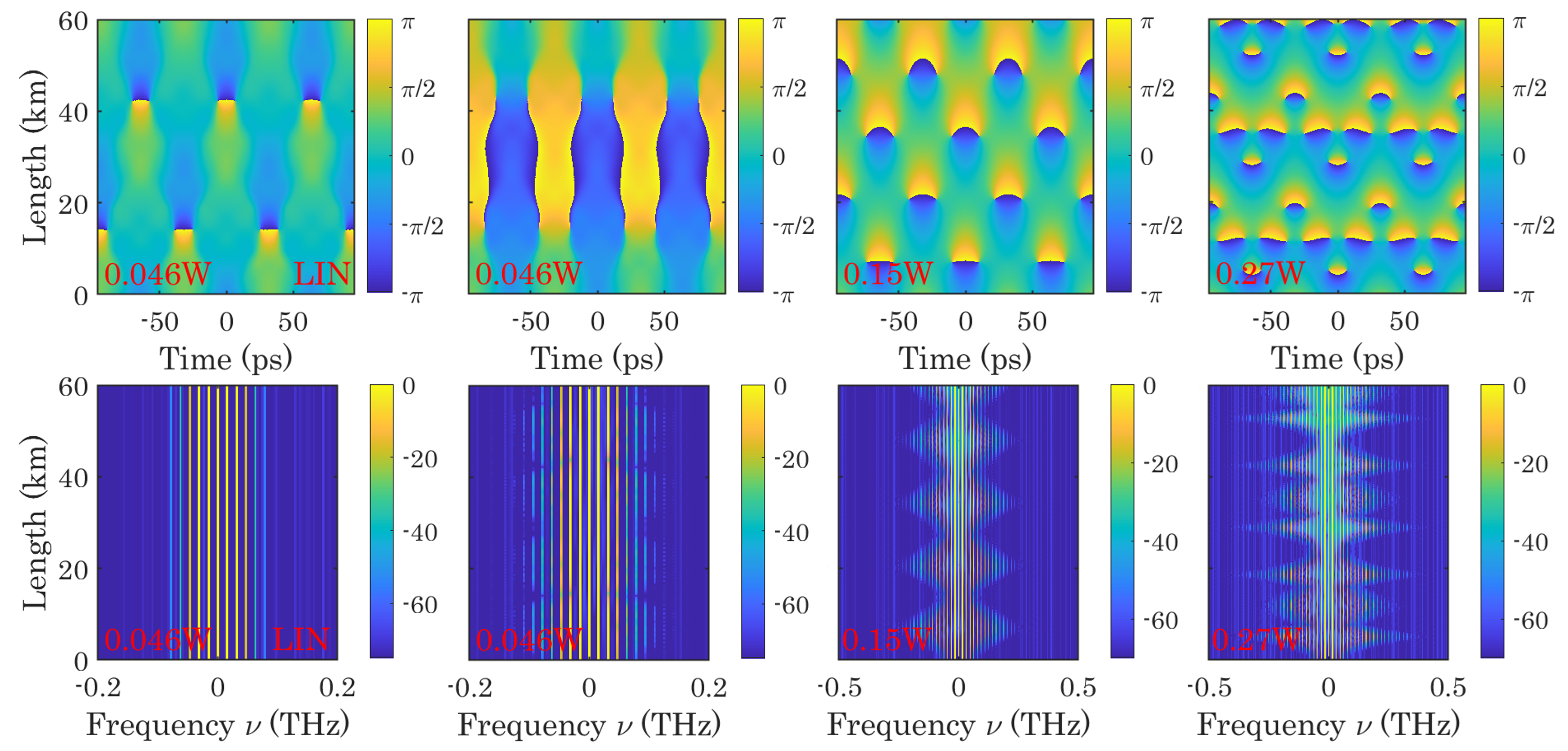}}
\caption{\textit{Top:} Phase evolution at regime-transition input powers of $P_{0}= 0.046~\text{W},$ $0.15~\text{W},$ and $0.27~\text{W}.$ \textit{Bottom:} Corresponding spectrum evolution. Left column with label LIN denotes the case for $\gamma = 0~\text{(W}\cdot\text{km)}^{-1}$ and corresponds to the \textit{linear} temporal Talbot Effect. The PM modulation depth is $m=1.$}
\label{fig:PHASE_SPEC}
\end{figure}

    Equipped with the knowledge gained in Secs.~\ref{sec:SRBA_Dispersion_Relation} and \ref{sec:SRBA_Effects}, I can now provide a theoretically backed-up explanation of different regimes and regime transitions of the linear and nonlinear temporal Talbot effect reported in Ref.~\cite{Zajnulina_2024} and presented in Fig.~\ref{fig:SRBA_Opt_Power}, left.  

\begin{figure}[htbp]
\centering
\fbox{\includegraphics[width=0.98\textwidth]{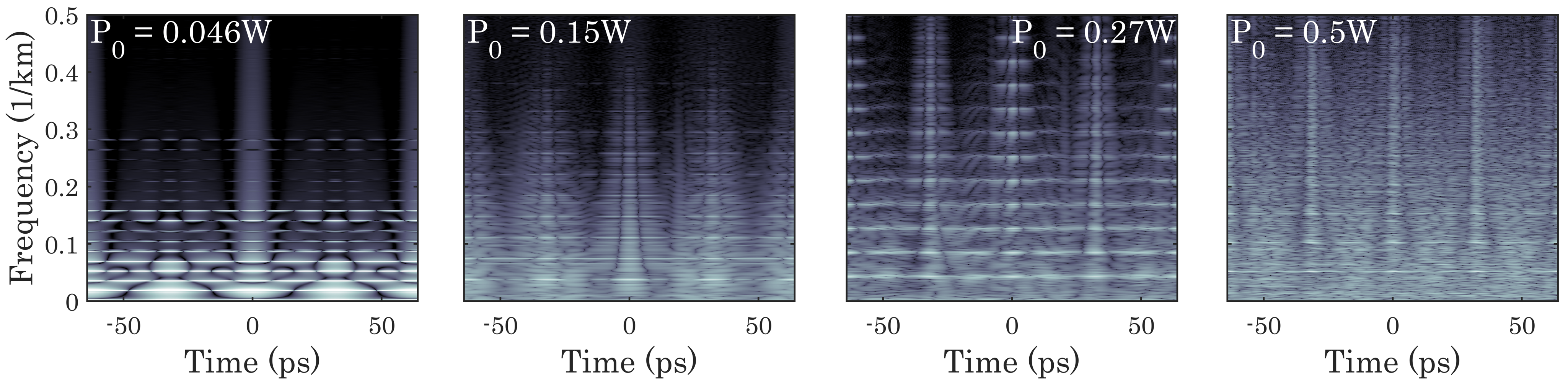}}
\caption{Time-resolved Soliton Radiation Beat Analysis (in dB) for input powers  $P_{0}= 0.046~\text{W},$ $0.15~\text{W},$ $0.27~\text{W},$ and and $0.5~\text{W}.$ The PM modulation depth is $m=1.$}
\label{fig:SRBA_over_Time}
\end{figure}

    \textbf{Regime I and the Harmonics of the Linear Temporal Talbot Effect for $P_{0}\rightarrow 0~\text{W}.$} Here, we see the emergence of ridges at higher harmonics and lower harmonics of the fundamental Talbot frequency $Z_{T}$ for $P_{0}\rightarrow 0~\text{W}.$ The Talbot frequency relates to the linear temporal Talbot effect and constitutes the inverse of the Talbot period \textcolor{black}{ $L_{T}=\frac{2\pi}{(2\pi\Omega)^2|\beta_{2}|}$} \cite{Zajnulina_2024}, i.e.
    \begin{equation}
      Z_{T}: =(2\pi\Omega)^{2}|\beta_{2}| = \frac{2\pi}{L_{T}} = 0.2217~\text{km}^{-1}.
      \label{equ:Z_Talbot}
  \end{equation}
    The seeding of the fundamental Talbot frequency as an SRBA ridge is explicable by the beating of the spatial mode $\zeta_{0}$ of the central frequency $\omega_{0} = 0~\text{THz}$ with the neighboring modes $\zeta_{\pm 1} = \frac{\beta_2}{2}(2\pi\Omega)^{2}$ for $P_{0}\rightarrow 0~\text{W}$ (Eq.~\ref{equ_DispersionRalation_Full}). 

    The appearance of higher and lower harmonics of $Z_{T}$ was already reported, but not exhaustively explained in Ref.~\cite{Zajnulina_2024}. To understand these harmonics, let us consider two examples of Eq.~\ref{equ:Power_Beating} using only a few comb lines. Thus, a pair of lines $\omega_{k}$ and $\omega_{l}$ would produce the following beating pattern:
   \begin{equation}
   \begin{aligned}      
       P(z) = &P_{0}|i^{k}J_{k}e^{i\zeta_{k}z} + i^{l}J_{l}e^{i\zeta_{l}z}|^{2} = \\=&P_{0}\left(J^{2}_{k} + J_{l}^{2} + 2J_{k}J_{l}\cos{\left([\zeta_{k}-\zeta_{l}]z + \frac{\pi}{2}[k-l]\right)}\right). \end{aligned}   
   \label{equ:interference_two_lines}
   \end{equation}
 For $P_{0}\rightarrow 0~\text{W},$ Eq.~\ref{equ:interference_two_lines} simplifies to:
    \begin{equation}
        P(z) =  P_{0}\left(J^{2}_{k} + J_{l}^{2} + 2J_{k}J_{l}\cos{\left(\frac{Z_{T}}{2}[k^{2} - l^{2}]z + \frac{\pi}{2}[k-l]\right)}\right),
   \end{equation}
   which shows that a pairwise beating between the lines with high values of the indices $k$ and $l$ can generate higher harmonics at multiples of $Z_{T}/2.$ 
   
   Three lines $\omega_{k},$ $\omega_{l},$ and $\omega_{m}$ would be involved in a more complex (higher-order) interference:
\begin{equation}
\begin{aligned}
    P(z) = P_{0}|i^{k}J_{k}e^{i\zeta_{k}z} + i^{l}J_{l}e^{i\zeta_{l}z} + i^{m}J_{m}e^{i\zeta_{m}z}|^{2} =\\
    = P_{0}(J^{2}_{k} + J_{l}^{2} + J_{m}^{2} +\\ 2J_{k}J_{l}\cos{([\zeta_{k}-\zeta_{l}]z + \phi_{kl})} +\\ 2J_{l}J_{m}\cos{([\zeta_{l}-\zeta_{m}]z+\phi_{lm})} +\\ 2J_{k}J_{m}\cos{([\zeta_{k}-\zeta_{m}]z + \phi_{km})})
\end{aligned}
\label{equ:interference_three_lines}
\end{equation}
  with an overlap of cosine functions of different periods. For $P_{0}\rightarrow 0~\text{W}$, these periods are proportional to $\frac{Z_{T}}{2}[k^{2} - l^{2}],$ $\frac{Z_{T}}{2}[l^{2} - m^{2}],$ and $\frac{Z_{T}}{2}[k^{2} - m^{2}]$. Depending on the line-indices values, constructive overlap, i.e., interference between these cosines, can happen at longer propagation distances, resulting in subharmonics of the SRBA plot. Phase factors $\phi_{ij}:= \frac{\pi}{2}[i-j]$ play a role in achieving constructive interference of spatial modes or preventing it.    
  
  These two simple examples of pairwise and 3-line (higher-order) interference show that the optical power $P(z)$ oscillates with fractional multiples of $Z_{T}/2$ for $P_{0}\rightarrow 0~\text{W}$. The effect behind is the \textit{linear} temporal Talbot effect \cite{Jannson_1981}, \cite{Azana_2003}, \cite{Cortes_2019}. This interference translates to the appearance of higher and lower harmonics of $Z_{T}$ in the SRBA plot for $P_{0}\rightarrow 0~\text{W}.$ As the richest and most interesting ridges arise at subharmonics of $Z_{T},$ I conclude that it is the higher-order interference (or beating) between several comb lines that sets the tone for the dynamics of a phase-modulated CW laser field  (Fig.~\ref{fig:SRBA_Opt_Power}, left). In other words, the \textit{linear} temporal Talbot effect seeds the spatio-temporal distribution of optical pulses through interference of field's spatial modes. With increasing input power and non-negligible nonlinear effects, the spatio-temporal distribution will be modified according to Eq.~\ref{equ_DispersionRalation_Full}, connecting temporal Talbot effect with FPUT phenomena in optical fibers.
  
  The strong presence of higher-order interference inherited from the linear temporal Talbot effect and including both the optical pulses and the background on which they sit is seen in Fig.~\ref{fig:SRBA_over_Time} for $P_{0}= 0.046~\text{W}$. It is recognizable by horizontal lines of fixed SRBA spatial frequency $<Z_{T}$ stretching over the whole temporal window.
     
\begin{figure}[htbp]
\centering
\fbox{\includegraphics[width=0.65\textwidth]{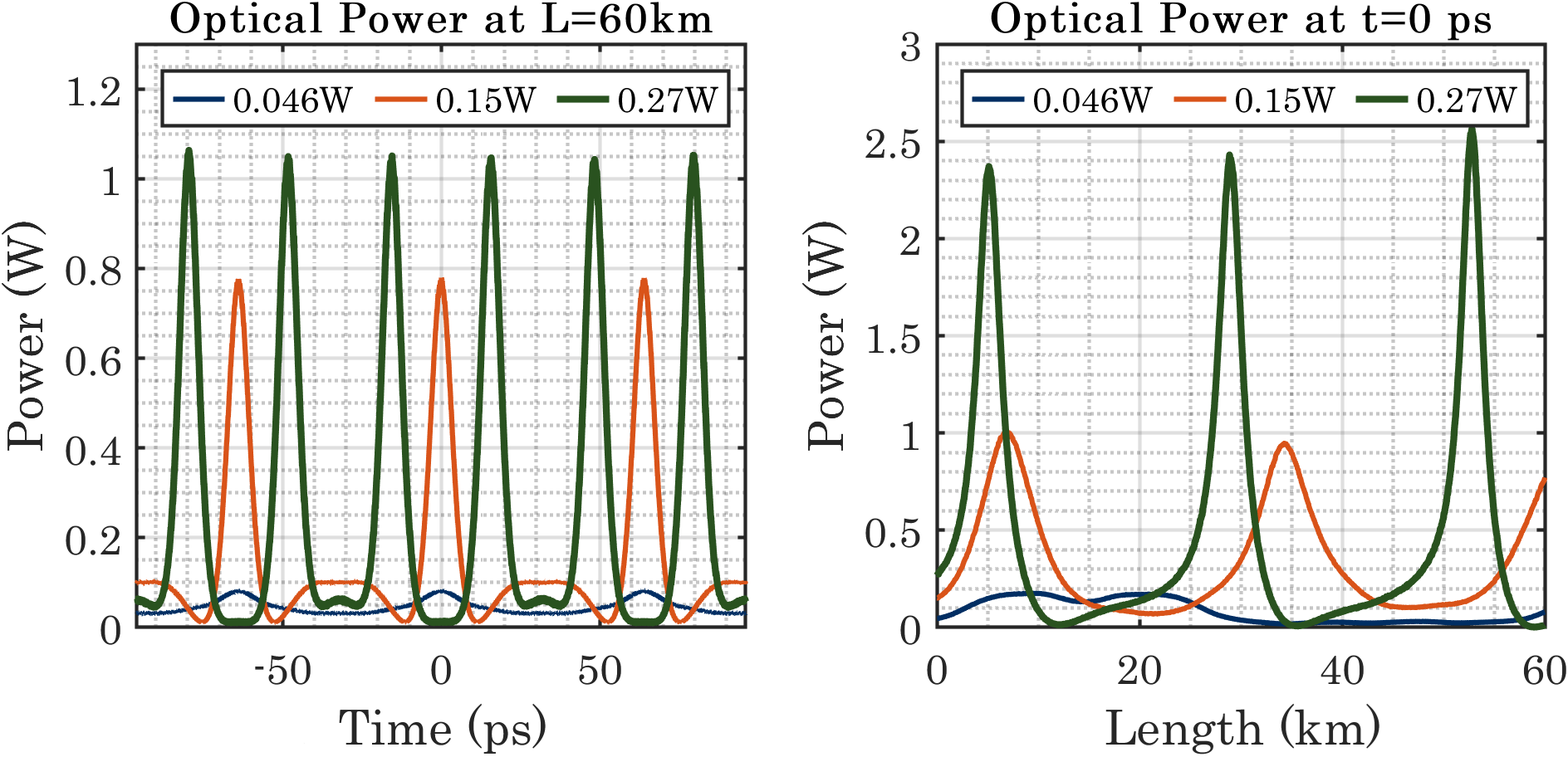}}
\caption{\textit{Left:} Temporal evolution of the optical power at fiber length of $L=60~\text{km}$ for input powers of $P_{0}= 0.046~\text{W},$ $0.15~\text{W},$ and $0.27~\text{W}.$ \textit{Right:} Corresponding evolution along the propagation distance $z$ at the center of the temporal window $t=0~\text{ps}.$ The PM modulation depth is $m=1.$}
\label{fig:Opt_Power_Distance_Time_PHASE}
\end{figure}

    \textbf{Regime I to Regime II Transition at $P_{0}= 0.046~\text{W}$.} With increasing input power, the nonlinear terms proportional to $\gamma P_{0}$ start counterbalancing the GVD term in the spatial modes $\zeta_{k}$  of Eq.~\ref{equ_DispersionRalation_Full}. These terms become non-negligible at $P_{0}= 0.046~\text{W},$ the transition point from the \textit{linear} to \textit{nonlinear} temporal Talbot effect (Fig.~\ref{fig:SRBA_Opt_Power}, left). 

    The existence of this transition can be explained from the perspective of soliton theory. Thus, fundamental solitons exist for soliton orders of $0.5\leq N < 1.5$ (\cite{Taylor_2009}, \cite{Boehm_2006}). It implies a minimal input power that needs to be provided to the system for a fundamental soliton to be created. This input power corresponds to the order $N=0.5$ and is exactly $P_{0}=0.046~\text{W}$ following the relationship (\cite{Zajnulina_2015}, \cite{Zajnulina_2017}, \cite{Zajnulina_2025}):
\begin{equation}
    N^2 := \frac{L_{\text{D}}}{L_{\text{NL}}} = \frac{\gamma P_{0}}{(2\pi\Omega)^2 |\beta_{2}|}
    \label{equ:Soliton_Order_N}
\end{equation}
    with dispersion length $L_{\text{D}}:=\frac{1}{(2\pi\Omega)^2|\beta_{2}|}= \frac{L_{T}}{2\pi}$ and nonlinear length $L_{\text{NL}}: = \frac{1}{\gamma P_{0}}.$ 

    The impact of the nonlinearity becomes visible in the MI-related breathing of the optical spectrum in Fig.~\ref{fig:PHASE_SPEC}, where you can compare the spectrum for $P_{0} = 0.046~\text{W}$ with a spectrum of the linear case (LIN) for which the nonlinearity was set to zero. Two small homoclinic loops in the field trajectory reveal first solitary waves along the propagation distance (\cite{Balmforth_1994}, \cite{Bi_2006}, \cite{Bandara_2023}, \cite{Alzaleq_2024}), although they are not recognizable as such in the optical-power evolution (Fig.~\ref{fig:Trajectories}, \cite{Zajnulina_2024}). In the SRBA representation (Fig.~\ref{fig:SRBA_Opt_Power}), this $P_{0}$ transition point marks the appearance of the input-power-dependent strong curvature of the ridges.    

    Initially missed in the foundational Ref.~\cite{Zajnulina_2024}, the input power $P_{0}=0.046~\text{W}$ constitutes a fundamentally important point where the Kerr nonlinearity of the fiber becomes non-negligible, thus, separating the \textit{linear} temporal Talbot effect from the \textit{nonlinear} Talbot effect.

    \textbf{Regime II of A-Type Breathers, $0.046~\text{W} < P_{0}\leq 0.15~\text{W}.$} A comparison of phase and spectrum in Fig.~\ref{fig:PHASE_SPEC} and trajectories in Fig.~\ref{fig:Trajectories} for $P_{0}=0.046~\text{W}$ and $0.15~\text{W}$ shows a strong resemblance with the theoretical and experimental reports of doubly-periodic A-type breather solutions of the NLS (Eq.~\ref{equ:NLS}) \cite{Kimmoun_2016}, \cite{COnforti_2020}, \cite{Vanderhaegen_2020a}, \cite{Vanderhaegen_2021}, \cite{Schiek_2021}.  To observe A-type breathers experimentally, the authors of Refs.~\cite{Vanderhaegen_2020a} and \cite{Vanderhaegen_2021} deploy a bidirectionally pumped SMF to compensate for the fiber losses and use two CW lasers to seed a frequency comb that has only a few lines as an initial condition. Their experimental setup effectively replicates the system I study here theoretically and numerically (Fig.~\ref{fig:SETUP}). From the similarity of the systems and reported results, I conclude that Regime II is occupied by A-type breathers \cite{COnforti_2020}. 

    Whereas the authors of Ref.~\cite{Vanderhaegen_2021} impose a phase shift of $-\frac{\pi}{2}$ between the pump and the signal waves in their 3-wave experiment to generate doubly-periodic A-type breather solutions of the NLS, such a phase shift is naturally introduced by the linear temporal Talbot effect in a phase-modulated CW field propagating through an SMF \cite{Jannson_1981}, \cite{Azana_2003}. 

    As seen in Fig.~\ref{fig:SRBA_Opt_Power}, left, A-type breather solutions are recognizable by pitchfork or rake-shaped SRBA ridge bundles. The most intense A-type breathers are seeded by subharmonics of the fundamental Talbot frequency $Z_{T},$ indicating higher-order interference between the comb lines and the corresponding spatial modes (Eqs.~\ref{equ_DispersionRalation_Full}, \ref{equ:interference_three_lines}) as a decisive mechanism for the formation of this type of NLS solutions. 

    For input powers $P_{0}\geq 0.046~\text{W},$ both effects, FWM and (indirectly) XPM, contribute to the broadening of the optical spectrum. This contribution manifests itself in a strong curvature input-power-dependent pulling together of the SRBA ridges (Fig.~\ref{fig:SRBA_Opt_Power}, left). However, only the comb lines that are close to the central line $\omega_{0} = 0~\text{THz}$ are intense enough to support these effects (Eq.~\ref{equ_DispersionRalation_Full}). For FWM, this implies that it is rather \textit{regular} FWM that takes place in this regime than the cascaded one (cf. \cite{COnforti_2020}, \cite{Vanderhaegen_2020a}, \cite{Vanderhaegen_2021}). 

    The spectral broadening due to \textit{regular} FWM leads to the temporal compression and spatio-temporal localization of the optical pulses on a background, i.e., A-type breathers (Fig.~\ref{fig:Opt_Power_Distance_Time_PHASE}). This results in granulation of the optical phase (Fig.~\ref{fig:PHASE_SPEC}) for $P_{0}= 0.15~\text{W}.$ Temporal localization is also well seen in Fig.~\ref{fig:SRBA_over_Time}  for $P_{0}=0.15~\text{W}$ as compared to the case of $P_{0}=0.046~\text{W}.$ Here, we see the appearance of horizontal SRBA ridges related to solitons at the locations of temporal pulse localization, indicating that the A-type breather's background is hardly involved in overall nonlinear dynamics.  

    \textbf{Regime III of Soliton Crystals, $0.15~\text{W}<P_{0}\leq 0.27~\text{W}.$} As the spectrum broadens and FWM locks spatial-mode phases concentrating energy into fewer dominant lines, the relative XPM contribution to curvature of SRBA ridges diminishes, allowing SPM and FWM to become more dominant nonlinear effects for $P_{0}=0.15~\text{W}$ (Eq.~\ref{equ_DispersionRalation_Full}, Fig.~\ref{fig:PHASE_SPEC}). 

    The trajectories in Fig.~\ref{fig:Trajectories} indicate a significant input-power-dependent transformation of the optical field in regime III. Thus, the trajectories of A-type breathers (\cite{COnforti_2020}, \cite{Vanderhaegen_2020a}, \cite{Vanderhaegen_2020b}) transfer to periodic solitons over the propagation distance $z,$ the latter recognizable by a homoclinic loop embedded in a circular oscillatory orbit at $P_{0}=0.27~\text{W}.$ The optical phase is highly, but regularly granulated for $P_{0}=0.27~\text{W}$ (Fig.~\ref{fig:PHASE_SPEC}). 

    In Fig.~\ref{fig:Opt_Power_Distance_Time_PHASE}, $P_{0}= 0.27~\text{W}$ we see that solitons that evolved in regime III do not match the description of Peregrine solitons, as the latter sit on a homogeneous background (\cite{Kibler_2010}, \cite{Hammani_2011}, \cite{Tikan_2017}, \cite{Ye_2020}, \cite{Karjanto_2021}). Here, however, the spatio-temporal localization goes almost down to zero, constituting soliton crystals as a spatio-temporally periodic soliton compounds and rather a separate class of NLS solutions (\cite{Zajnulina_2015}, \cite{Zajnulina_2017}). The regular spatio-temporal order is seeded by the higher-order interference of the temporal Talbot effect (cf. Eq.~\ref{equ:interference_three_lines}).    

    Comparison of the plots for $P_{0}= 0.15~\text{W}$ and $P_{0}= 0.27~\text{W}$ in Fig.~\ref{fig:SRBA_over_Time} reveals stronger temporal localization of the pulses than in regime II. The well-pronounced and almost sharp intensity of the soliton-related SRBA harmonics below and above the fundamental Talbot frequency $Z_{T}$ indicates strong phase locking of the spatial field modes for $P_{0} = 0.27~\text{W}$, which is the result of FWM (Eq.~\ref{equ_DispersionRalation_Full}). 

    The richness of harmonics above $Z_{T}$ is attributable to the beating of lines with high indices (cf. Eqs.~\ref{equ:interference_two_lines}, \ref{equ:interference_three_lines}). Such frequencies are the result of \text{cascaded} FWM. In other words, soliton crystals in single-pass systems (\cite{Zajnulina_2015}, \cite{Zajnulina_2017}, \cite{Zajnulina_2024}) result from cascaded FWM, contrary to A-type breathers that originate from regular FWM. 

    As soliton crystals in single-pass optical fibers require comparably high input powers (Fig.~\ref{fig:SRBA_Opt_Power}, left) and are associated with cascaded FWM, they are no different from soliton crystals in cavities (cf. \cite{Cole_2017}, \cite{Singh_2024}, \cite{Lu_2021}, \cite{Karpov_2019}) allowing for a more generalized approach in the nonlinear-wave evolution in single-pass and cavity systems.
  
\textbf{Regime IV of Separated Solitons and Soliton Gas, $P_{0}>0.27~\text{W}.$}
  For input powers $P_{0}>0.27~\text{W},$ the SPM becomes dominant, which is recognizable by comb-line broadening in the corresponding Fig.~\ref{fig:PHASE_SPEC}. It implies that the dynamics of the spatial modes are driven by the balance between GVD and SPM, a characteristic of soliton evolution (\cite{Agrawal_2019}, \cite{Taylor_2009}):
  
\begin{equation}
    \zeta_{k}(\omega_{k})\rightarrow\frac{\beta_{2}}{2}\omega_{k}^{2} + \gamma P_{0}J_{k}^{2}.
    \label{equ:Z_soliton}
\end{equation}

   At $P_{0}= 0.27~\text{W}$, the spatial modes are (still) well phase-locked due to FWM of regime III, but significantly dephase due to SPM with increasing input power (Fig.~\ref{fig:SRBA_over_Time}, compare $P_{0}=0.27~\text{W}$ and $P_{0}=0.5~\text{W}).$  As a result, the rigidness of the spatio-temporal distribution initially imprinted by the Talbot effect dissolves, giving stage to the development of \textit{soliton gas,} a state of multiple weakly interacting solitons ( ~\cite{Suret_2024}). Their irregular spatio-temporal distribution is reflected in the embedding of several homoclinic orbits within each other (Fig.~\ref{fig:Trajectories}, $P_{0}=0.5~\text{W}$).
   
   The soliton gas consists not only of solitons that result from the dissolution of the soliton crystals at $P_{0}= 0.27~\text{W},$ but also of new-born ones as seen at $Z = 0~\text{km}^{-1}$ and $P_{0}= 0.4~\text{W}$ in Fig.~\ref{fig:SRBA_Opt_Power}, left (\cite{Boehm_2006}, \cite{Mitschke_2017a}, \cite{Zajnulina_2015}, \cite{Zajnulina_2017}). 

    To conclude, i) the initial spatio-temporal distribution of the optical pulses is seeded by (higher-order) interference of the spatial modes, which is due to the \textit{linear} temporal Talbot effect; ii) the Kerr nonlinearity of the fiber and its effects modify this spatio-temporal distribution and can be integrated in the framework of the \textit{nonlinear} temporal Talbot effect by considering frequency-dependent spatial modes. Regime transitions of the \textit{nonlinear} temporal Talbot effect are governed by a shifting balance between GVD, SPM, XPM, and FWM in the spatial-mode dispersion relation. At low powers, higher-order interference from the linear Talbot effect dominates, while increasing input power drives spectral broadening, spatial-mode phase locking, and spatio-temporal pulse localization through regular and cascaded FWM. Regular FWM leads to the formation of A-type breathers, whereas cascaded FWM involves the generation of soliton crystals. At the highest powers, SPM prevails, breaking spatial-mode phase locking and leading to a soliton gas of weakly interacting pulses.

\subsection{Soliton Encoding by Frequency Comb Lines}
\label{sec:Lines_Solitons}
    The dispersion relation Eq.~\ref{equ_DispersionRalation_Full} and findings of Secs.~\ref{sec:SRBA_Dispersion_Relation} and \ref{sec:Transitions} explain well not only the input-power-dependent regimes of the optical-power evolution in the context of the nonlinear temporal Talbot effect, but they also explain the phenomenon that was first observed and reported in Ref.~\cite{Zajnulina_2024}. Namely, if we consider each line of a frequency comb separately and perform SRBA on it by calculating its spatial frequencies, we will see encoding of some line-specific solitons as a balance between the GVD and SPM in the corresponding spatial mode $\zeta_{k}$ (Eq.~\ref{equ:Z_soliton}). At the same time, we will observe some solitons that are common throughout the comb and can be attributed to collective effects such as FWM. 

\begin{figure}[htbp]
\centering
\fbox{\includegraphics[width=0.85\textwidth]{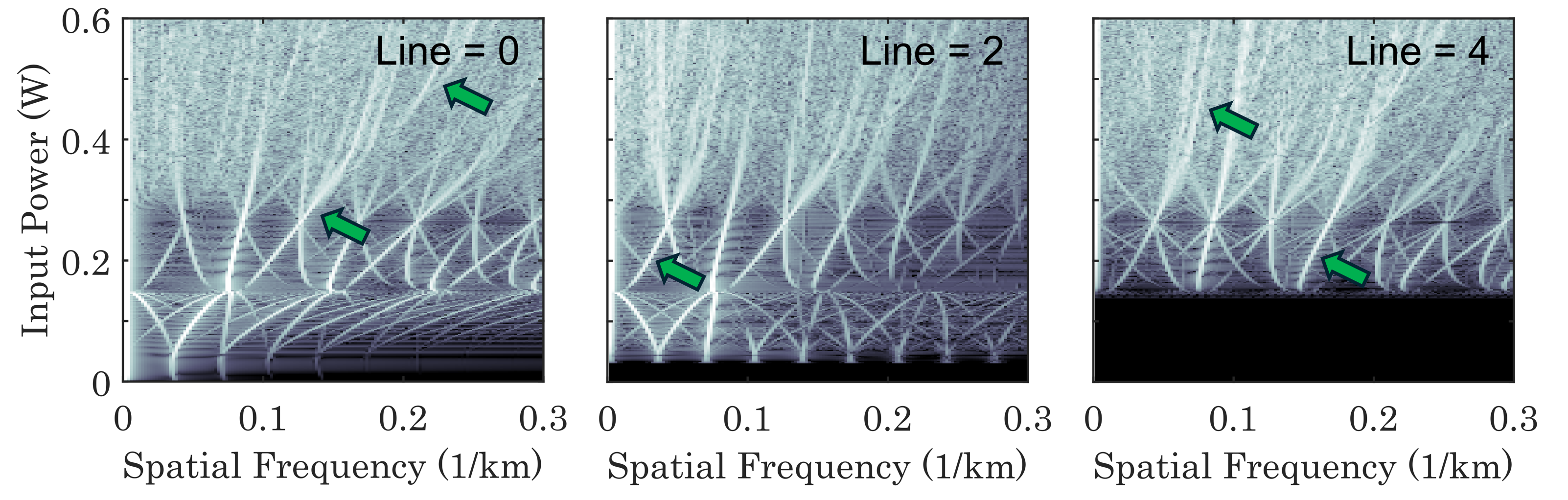}}
\caption{Soliton Radiation Beat Analysis (in dB) of input-power dependent regimes for the central frequency-comb line (Line = 0) at $\omega_{0}$ and lines with indices $k= 2$ and $k = 4$ (Line = 2 and Line = 4). The PM modulation depth is $m=1$ \cite{Zajnulina_2024}.}
\label{fig:SRBA_Lines}
\end{figure}

    Fig.~\ref{fig:SRBA_Lines} shows an example for the SRBA of the central comb line and lines with indices $k=2$ and $k=4.$ In this figure, green arrows indicate the most intense line-specific solitonic ridges. Additionally, we observe black stripes whose input-power-dependent region increases with the index of the line. In the framework of SRBA, black stripes indicate that there is no oscillation over the propagation distance for the given line and input power; the amplitude of the line is too weak to participate in the nonlinear interaction with other lines via FWM and XPM. We also see that the regime of A-type breathers is covered by lines that are close to the central line (Line = 2). This observation supports the hypothesis of \textit{regular} FWM involving only a few lines in this regime (Sec.~\ref{sec:Transitions}). The regime of soliton crystals needs a higher number of comb lines to be set off (Line = 4). Therefore, the black stripe is broader. This observation supports the hypothesis of \textit{cascaded} FWM being primarily involved in the formation of these waves.

    In short, SRBA of individual frequency-comb lines shows that low-index lines near the central frequency support line-specific solitons and A-type breathers via regular FWM, while higher-index lines enable soliton crystals through cascaded FWM. 

\subsection{Effect of the Phase-Modulator Modulation Depth}
\label{sec:Modulation_Depth}
\begin{figure}[htbp]
\centering
\fbox{\includegraphics[width=0.95\textwidth]{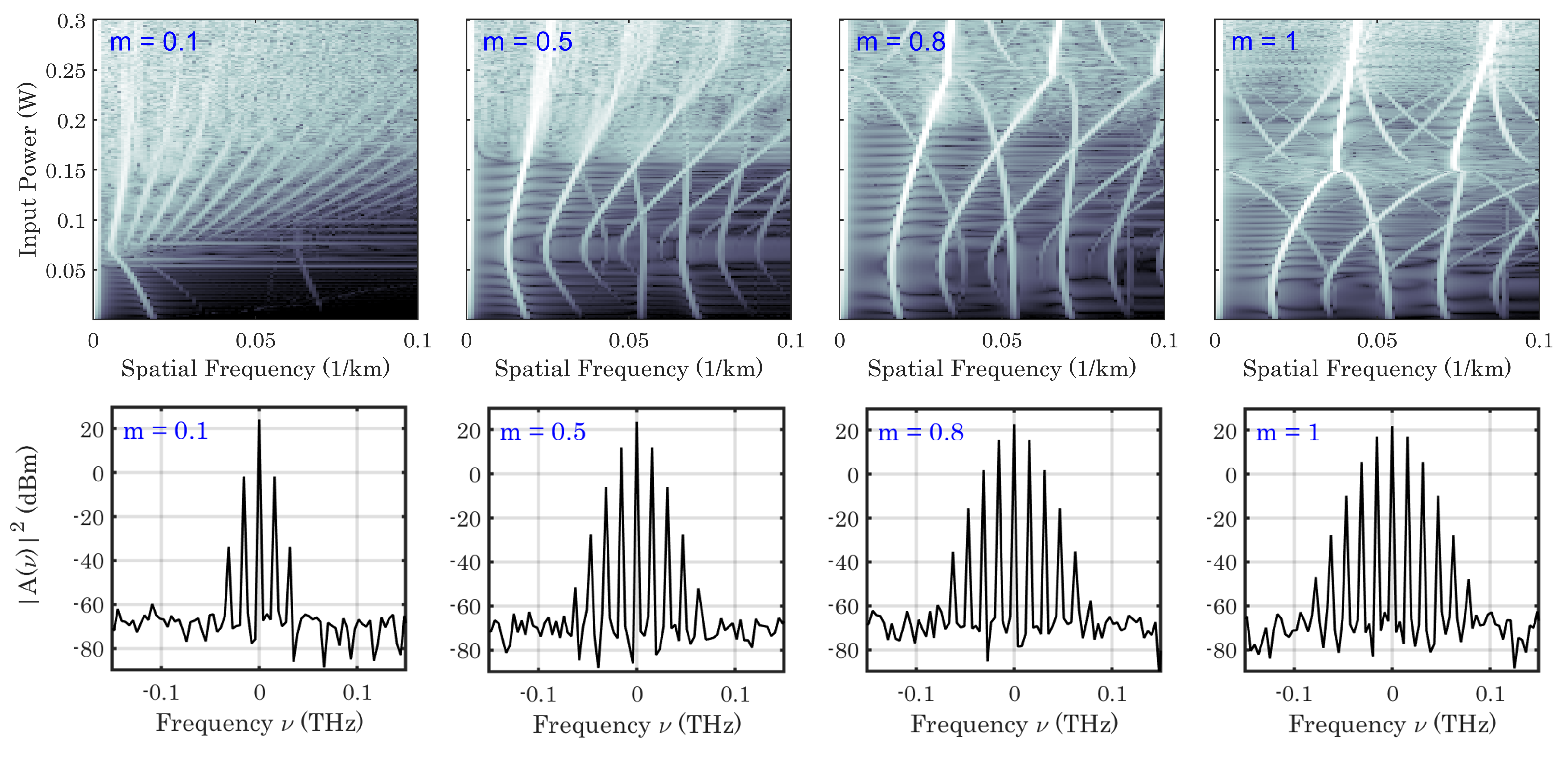}}
\caption{\textit{Top:} Soliton Radiation Beat Analysis (in dB) of input-power dependent regimes for modulation depths of $\text{m}= 0.1, 0.5, 0.8, 1.$ \textit{Bottom:} Corresponding initial frequency combs at input power of $P_{0}=0.27~\text{W}.$}
\label{fig:SRBA_ModDepthDiff_Below1}
\end{figure}

    Let us now study the impact of the PM modulation depth $m$ on the development of the nonlinear temporal Talbot effect regimes. \textcolor{black}{According to the Jacobi-Anger expansion of the initial condition, the parameter $m$ directly controls the number and relative amplitudes of frequency-comb lines through the Bessel-function coefficients $J_k(m)$ (Eqs.~\ref{equ:IC}, \ref{equ_IC_JA}). Thus, higher values of $m$ yield an initial comb with more lines. The aim of this section is to show that the PM modulation depth is an additional parameter in the dynamics governed by the nonlinear temporal Talbot effect. Other system paramters remain the same as in Sec.~\ref{sec:In_Advance_Discussion}.} 
    
    Fig.~\ref{fig:SRBA_ModDepthDiff_Below1}, top, shows SRBA plots for different values of modulation depth, ranging from $m=0.1$ (weak modulation) to $m=1$ (quite strong modulation with dynamics studied in previous sections). The corresponding bottom panel shows how the number of initial comb lines increases with the value of $m.$ 
    
    As seen in the top panel of Fig.~\ref{fig:SRBA_ModDepthDiff_Below1}, the dynamics changes significantly with the value of $m,$ starting with a case of only two distinguishable regimes $(m= 0.1)$ and continuously going to a case where we can see A-type breathers, and soliton crystals $(m= 1,$ Sec.~\ref{sec:Transitions}). Apparently, the number and the type of different regimes relate to the number of initial comb lines (Fig.~\ref{fig:SRBA_ModDepthDiff_Below1}, bottom). 
    
    As shown in the next step, this observation supports the discussions in Sections~\ref{sec:SRBA_Dispersion_Relation} and \ref{sec:SRBA_Effects}, particularly the discussion about the impact of the type of FWM on the dynamics. That is, \textit{regular} FWM with a low number of comb lines involved (initial and generated ones) leads to A-type breathers, whereas \textit{cascaded} FWM yields soliton crystals. Considering the optical-power evolution and the trajectories for different modulation depths $m$ will help us understand different regimes in Fig.~\ref{fig:SRBA_ModDepthDiff_Below1}, top, and their evolution from one to another with increasing $m.$ 

\begin{figure}[htbp]
\centering
\fbox{\includegraphics[width=0.95\textwidth]{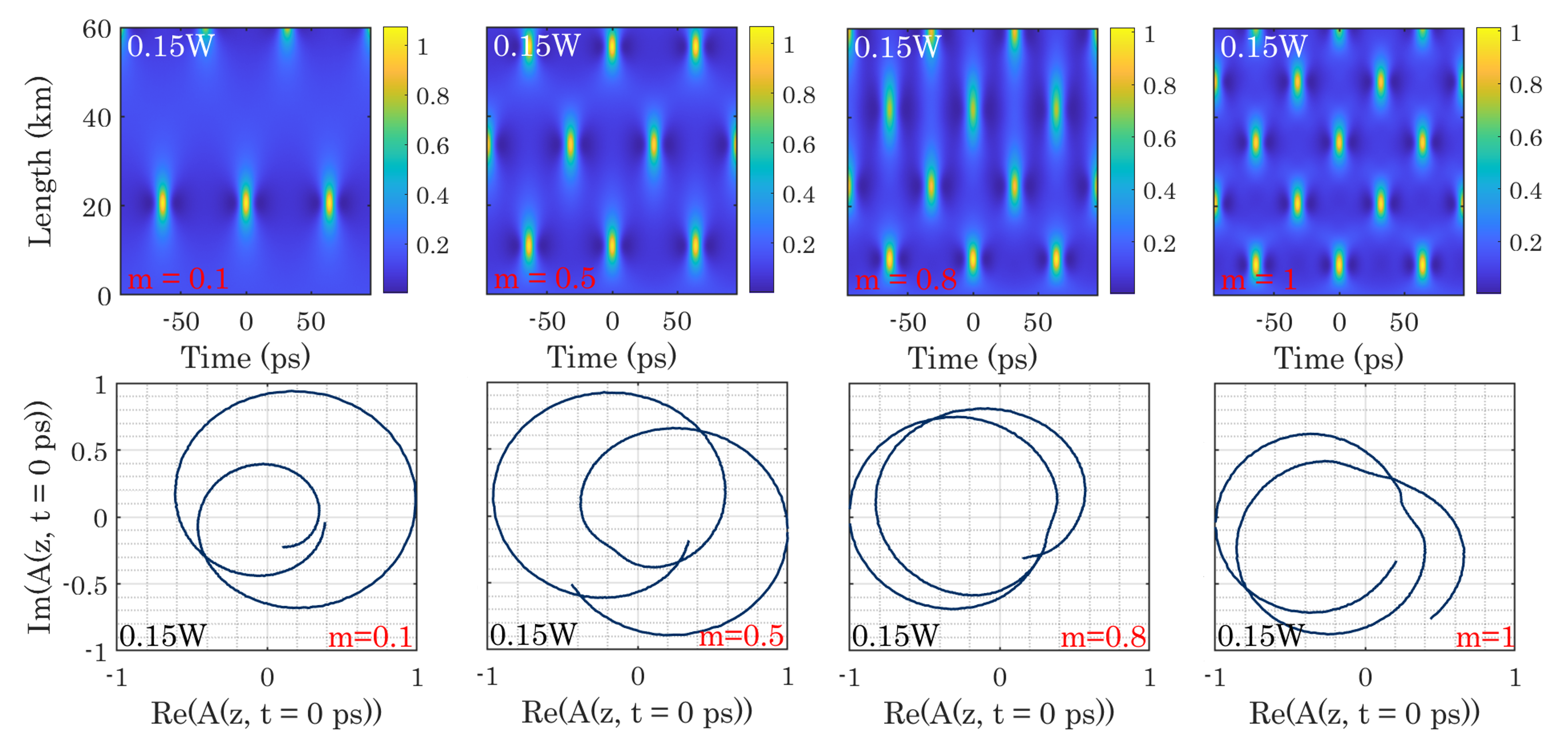}}
\caption{\textit{Top:} Optical power evolution (in W) along the fiber propagation distance $z$ for modulation depths of $\text{m}= 0.1, 0.5, 0.8, 1$ and input power of $P_{0}= 0.15~\text{W}.$  \textit{Bottom:} Corresponding trajectories for $t = 0~\text{ps},$ i.e., the center of the chosen temporal window.}
\label{fig:DiffMod_15}
\end{figure}

\begin{figure}[htbp]
\centering
\fbox{\includegraphics[width=0.95\textwidth]{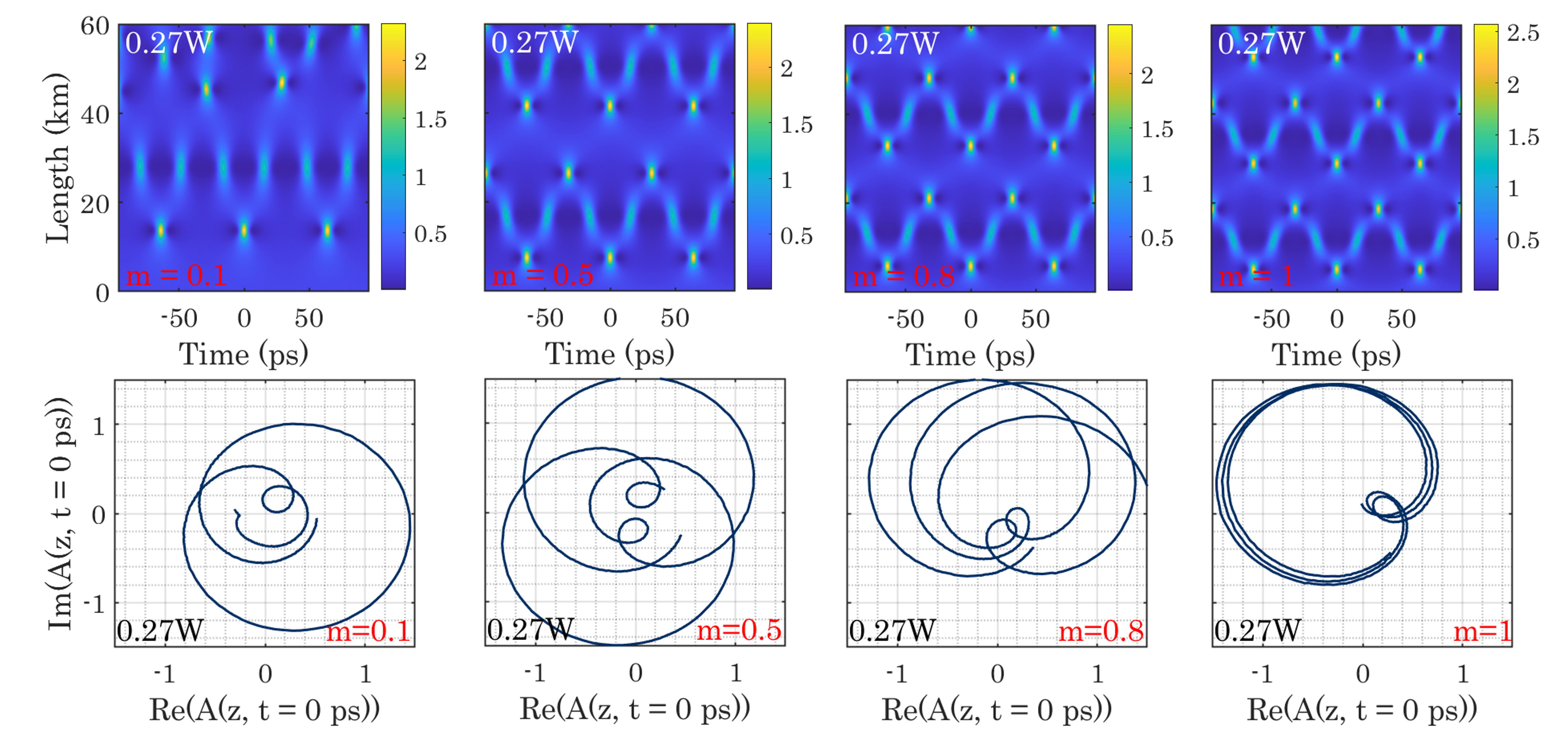}}
\caption{\textit{Top:} Optical power evolution (in W) along the fiber propagation distance $z$ for modulation depths of $\text{m}= 0.1, 0.5, 0.8, 1$ and input power of $P_{0}= 0.27~\text{W}.$  \textit{Bottom:} Corresponding trajectories for $t = 0~\text{ps},$ i.e., the center of the chosen temporal window.}
\label{fig:DiffMod_27}
\end{figure}

    Figs.~\ref{fig:DiffMod_15} and ~\ref{fig:DiffMod_27} show optical-power evolution and the corresponding trajectories for different values of the modulation depth $m$ for $P_{0}= 0.15~\text{W}$ and $P_{0}= 0.27~\text{W}.$

    For $P_{0}= 0.15~\text{W}$ and $m=0.1,$ the optical power shows a pulse structure of an AB under FPUT with $\pi-$shifted recurrence after a comparably long distance (cf. \cite{Kimmoun_2016}, \cite{Vanderhaegen_2020b}). The corresponding trajectory exhibits a homoclinic loop of a soliton embedded in a circular, periodic structure (cf.~\cite{Balmforth_1994}, \cite{Bi_2006}, \cite{Bandara_2023}, \cite{Alzaleq_2024}). The SRBA exhibits a fan of separated-soliton ridges and their overtones, driven primarily by SPM. With these characteristics, I conclude that we deal here with an FPUT AB that is close to its limit case of a train of Peregrine solitons (\cite{Kibler_2010}, \cite{Hammani_2011}, \cite{Tikan_2017}). The initial comb has only five lines (Fig.~\ref{fig:SRBA_ModDepthDiff_Below1}, bottom). Due to their limited number and relatively low input power, these lines engage in \textit{regular} FWM rather than the cascaded one. 

    An increasing PM modulation depth produces more initial comb lines, allowing XPM and FWM to contribute to the shift of the SRBA ridges away from the spatial frequency $Z=0~\text{km}^{-1}$ $(m=0.5)$ and building pitchforks of A-type breathers and fans of soliton crystals for $m=0.8$ and $m=1$ (Fig.~\ref{fig:SRBA_ModDepthDiff_Below1}, top). In general, shifting away from $Z=0~\text{km}^{-1}$ implies a decrease of the recurrence period well seen in Figs.~\ref{fig:DiffMod_15} and \ref{fig:DiffMod_27}.
 
    For $P_{0}= 0.15~\text{W}$ and increasing $m,$ the trajectories change the type of the NLS (Eq.~\ref{equ:NLS}) solution from an AB (close to a train of Peregrine solitons) to an A-type breather by enlarging the diameter of the homoclinic loop embedded in a circular, periodic structure and transforming the overall shape (\cite{COnforti_2020}). 

    For $P_{0}= 0.27~\text{W}$ and increasing $m,$ the trajectories are of a soliton type, embedding homoclinic loops in circular, periodic structures. The loops are, however, seemingly irregular for low modulation depths and gain order with increasing $m,$ resulting in a regularly periodic homoclinic loop for $m=1$ that was associated with a soliton crystal at input power of dissolution to separated solitons in Sec.~\ref{sec:Transitions}. The gain of order relates to the phase locking of optical field's spatial modes (Eq.~\ref{equ_DispersionRalation_Full}) due to FWM (Sec.~\ref{sec:Transitions}). With more comb lines involved and the development of rather \textit{cascaded} FWM as a result, the phase locking becomes stronger, facilitating the transition from a state that can be associated with an FPUT AB being close to Peregrine-soliton gas $(m=0.1)$ to a soliton crystal $(m=1).$ Again, through observing a gain of order in the trajectories, we see that many comb lines at higher input powers are involved in \textit{cascaded} FWM, resulting in soliton crystals. In contrast, \textit{regular} FWM at lower input powers yields A-type breathers.

\begin{figure}[htbp]
\centering
\fbox{\includegraphics[width=0.75\textwidth]{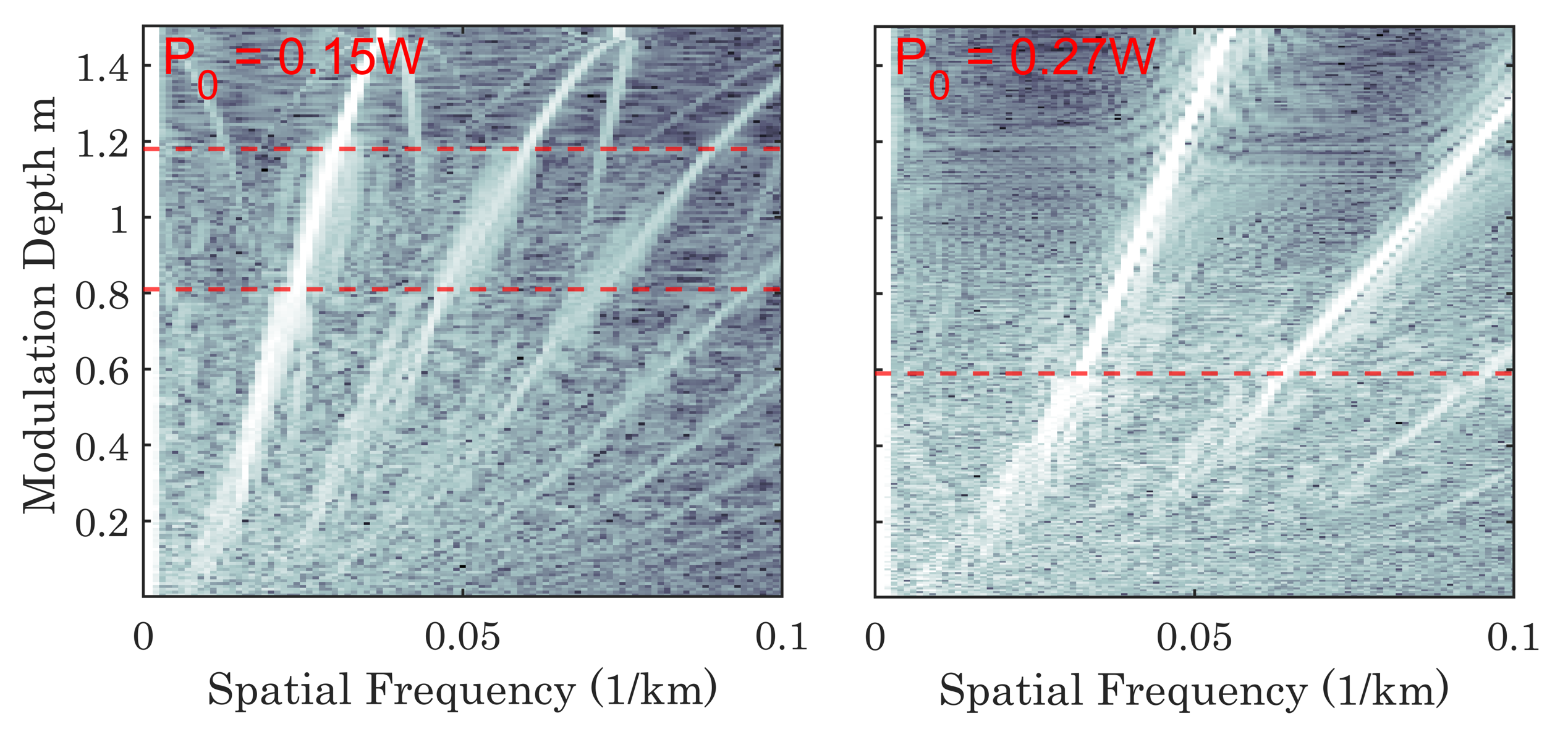}}
\caption{Soliton Radiation Beat Analysis (in dB) for different values of PM modulation depth $m$ and input powers of $P_{0}= 0.15~\text{W}$ and $0.27~\text{W}.$}
\label{fig:SRBA_ModDepth_HIGHRes}
\end{figure}

    When studying standard SRBA plots for fixed values of modulation depth (Fig.~\ref{fig:SRBA_ModDepthDiff_Below1}, top), one might gain the impression that the $m-$dependent transition between different types of the NLS solutions (Eq.~\ref{equ:NLS}) is smooth. However, changing the representation and studying SRBA plots for varying modulation depths at fixed input powers reveals the value of $m$ that can be associated with the NLS solution transitions (Fig.~\ref{fig:SRBA_ModDepth_HIGHRes}). Thus, we see a regime transition at $m=0.8$ for $P_{0}=0.15~\text{W}$ that can be associated with a change from an AB (being close to a train of Peregrine solitons) to an A-type breather. The modulation depth of $m= 0.6$ at $P_{0}=0.27~\text{W}$ denotes the transition from a gas of Peregrine solitons to a crystalline soliton structure.  

    To demonstrate that the dynamics is governed by separated (Peregrine) solitons rather than by collective states of breathers and crystals for low values of $m,$ let us consider the limit case of a weak modulation $m<<1.$ A Bessel function of the first kind that describes the comb-line amplitude with index $n$ (Eqs.~\ref{equ:IC}, \ref{equ:NLS_ansatz}) can be written as 
\begin{equation}
    J_n(m) = \sum_{k=0}^{\infty} \frac{(-1)^k}{k!\,(n+k)!} \left( \frac{m}{2} \right)^{2k+n}
\end{equation}
for positive integer orders $n\geq0.$ Taylor expansion of this function for $m<<1$ gives us following relationships for the central line at $\omega_{0}$ and the lines at $k=1$ and $k=2$ next to it (cf. \cite{Bender_2003}):
\begin{equation}
    J_{0}(m)\approx 1, \quad J_{1}(m)\approx \frac{m}{2}, \quad J_{2}(m)\approx\frac{m^{2}}{8}.    
\end{equation}
    This gives us the dispersion relation for the spatial modes (Eq.~\ref{equ_DispersionRalation_Full}):

\begin{align}
    \zeta_{0}(\omega_{0})&\approx& \frac{\beta_{2}}{2}\omega_{0}^{2}+\gamma P_{0}\left(1+\frac{m^{2}}{2}\right) + \mathcal{O}(m^{4}), \\
    \zeta_{1}(\omega_{1})&\approx& \frac{\beta_{2}}{2}(\omega_{0}+\Omega)^{2} + 2\gamma P_{0}\left(1+\frac{m^{2}}{2}\right)+\mathcal{O}(m^{4}),\\
    \zeta_{2}(\omega_{2})&\approx& \frac{\beta_{2}}{2}(\omega_{0}+2\Omega)^{2} + 2\gamma P_{0}\left(1+\frac{m^{2}}{4}\right)+\mathcal{O}(m^{4}).
\end{align}
    Here, we see that spatial modes of the most central lines collapse to a relationship between the GVD and self-phase modulation for $m\rightarrow 0$ (cf. Eq.~\ref{equ:Z_soliton}), which is a characteristic of solitons (\cite{Agrawal_2019}, \cite{Taylor_2009}). The effects that build collective states, namely, FWM and to some degree XPM, are missing. It confirms the observation that weak modulation yields (Peregrine) soliton formation; collective states such as breathers and soliton crystals are not even possible. However, with increasing input power and the value of $m,$ the contributions of FWM and XPM become possible, leading to the formation of A-type breathers and soliton crystals (Figs.~\ref{fig:DiffMod_15}, \ref{fig:DiffMod_27}).  

    In conclusion, the PM modulation depth $m$ is an important parameter that controls the number of initial frequency comb lines (Eq.~\ref{equ:IC}). Together with the input power $P_{0},$ it decides about the type and the transition between different solutions of the NLS. In particular, higher values of $m$ implying a higher number of initial comb lines lead to the formation of a soliton crystal, whereas a low number of $m$ results rather in a breather under FPUT being close to trains of Peregrine solitons. 
    
    From the perspective of the temporal Talbot effect, an increasing number of initial comb-lines increases the order of interference of spatial field modes (Eq.~\ref{equ_DispersionRalation_Full}), introducing SRBA ridges at multiples of $Z_{T}/2$ (Sec.~\ref{sec:Transitions}). These (arising) ridges correspond to spatial oscillations over shorter FPUT recurrence periods. From this observation and observations made in Sec.~\ref{sec:Transitions}, I conclude that FPUT recurrence of breathers directly relates to the interference order induced by the temporal Talbot effect. More studies are certainly needed to better link FPUT recurrence with the temporal Talbot effect, thereby developing an exhaustive theory of the \textit{nonlinear} temporal Talbot effect.   

\section{Conclusion}
\label{sec:conclusion}

    A phase-modulated continuous-wave (CW) laser field has a frequency comb in its spectrum. It is deployed in a wide range of optical systems and applications including wave shaping \cite{Finot_2015}, LiDARs \cite{Zhang_2023}, fundamental studies of nonlinear light propagation \cite{Andral_2020}, \cite{Friquet_2013}, and as an information carrier in frequency-multiplexed optical computing \cite{Zajnulina_2025}, \cite{Zajnulina_2025_Kerr}. In a nonlinear dispersive medium of an optical fiber or semiconductor, it is subject to the temporal Talbot effect and modulational instability (MI) \cite{Dudley_2009}. The temporal Talbot effect denotes the self-imaging of a pulse train due to the dispersion and is, per se, linear \cite{Jannson_1981}. MI is a nonlinear effect and is associated with the dynamical growth and evolution of periodic perturbations on a CW background \cite{Dudley_2009}. Their mutual action, amounting to an exhaustive theory of the \textit{nonlinear} temporal Talbot effect, has been hardly studied and understood, despite a growing interest over the last two decades. 

   In Ref. ~\cite{Zajnulina_2024}, my colleague Michael B\"ohm and I reported a strong relationship between the temporal Talbot effect and MI-related nonlinear-wave generation from a phase-modulated CW laser field in optical fibers. The results were achieved using the numerical technique of the Soliton Radiation Beat Analysis \cite{Boehm_2006}. Thus, we reported input-power-dependent transitions from a quasi-linear regime to the nonlinear regimes of the nonlinear Talbot effect that include soliton crystals and separated solitons (also called soliton gas due to a weak interaction between individual pulses). The transitions occurred at well-defined input power values. To the best of my knowledge, Ref.~\cite{Zajnulina_2024} reports the first results that discuss the nonlinear temporal Talbot effect in such detail. However, more studies were needed to better explain the observations presented in Ref.~\cite{Zajnulina_2024} and to develop a more profound theory of the nonlinear temporal Talbot effect.   

  Here, I build upon the results of Ref.~\cite{Zajnulina_2024} and study the transitions of the temporal Talbot effect and their physical mechanisms using Soliton Radiation Beat Analysis and deriving a dispersion relation for frequency-comb lines. This approach allows me to incorporate nonlinear effects in the framework of the initially \textit{linear} temporal Talbot effect, paving the way for the development of a concise theory of the \textit{nonlinear} temporal Talbot effect. It offers a new perspective and understanding of nonlinear wave generation in dispersion media. Here, I concentrate on optical fibers, but the results are also transferable to semiconductor waveguides (cf. \cite{Wu_2022}, \cite{Wu_2023}).

The following was achieved: 
   \begin{enumerate}[i)]
       \item I discussed that the system parameters chosen for the studies here (Sec.~\ref{sec:Setup}) do not support the formation of Akhmediev breathers under Fermi-Pasta-Ulam-Tsingou recurrence within the framework of the standard Akhmediev-breather theory. Instead, I show that the observed nonlinear waves constitute input-power-dependent regimes of A-type breathers, soliton crystals, and separated solitons (also called soliton gas \cite{Suret_2024}). Those are the regimes of the \textit{nonlinear} temporal Talbot effect. All regimes are separated by well-defined input powers that give rise to the interpretation that the governing Nonlinear Schr\"odinger Equation changes the type of its solutions at these values.
       
       \item I provide a theoretically backed-up explanation of physical effects that drive the regimes and regime segmentation. Thus, the regimes of the nonlinear temporal Talbot effect are strongly influenced by the self-phase modulation (SPM) of the frequency-comb lines as well as their four-wave mixing (FWM) and cross-phase modulation (XPM).
       
       \item I discuss that the regime transitions relate to the changes in the hierarchy of the effects involved. Thus, A-type breathers are primarily driven by FWM and XPM, whereas the impact of XPM declines in the case of soliton crystals, allowing FWM to take center stage. For separated solitons, both FWM and XPM are minor to SPM.
       
       \item I provide an explanation and theoretical evidence of the difference between the A-type breathers and soliton crystals. Both are spatio-temporally periodic. However, A-type breathers involve a low number of initial frequency-comb lines and arise from \textit{regular} FWM at quite low input powers. In contrast, soliton crystals involving many initial frequency-comb lines are a product of \textit{cascaded} FWM at higher input powers. With that, soliton crystals in single-pass optical fibers are no different from soliton crystals in cavities and should be considered as an independent solution class of the Nonlinear Schr\"odinger Equations (cf. \cite{Cole_2017}, \cite{Singh_2024}, \cite{Lu_2021}, \cite{Karpov_2019}).
       
       \item Most importantly, I show that the temporal Talbot effect determines the spatio-temporal distribution of the optical field by driving the (higher-order) spatial-mode interference and, thus, constitutes an underlying effect for optical-field evolution in the linear and nonlinear regimes. In particular, the spatial-mode interference relates to the the Fermi-Pasta-Ulam-Tsingou recurrence of optical pulses, an observation that needs to be studies in more detail for an exhaustive theory of the \textit{nonlinear} temporal Talbot effect. 
   \end{enumerate}

    To sum up, the results of the presented study show that the formation of the nonlinear waves in optical fibers from a phase-modulated CW input should be considered within the framework of the temporal Talbot effect, rather than by discussing them from the perspective of the modulational instability only. The recognition that the temporal Talbot effect underlies the spatio-temporal evolution of the optical field is fundamental. A theoretically backed-up incorporation of SPM, FWM, and XPM in the framework of the temporal Talbot effect contributes to the development of an exhaustive theory of the \textit{nonlinear} temporal Talbot effect. 

    The differentiation between the A-type breathers and soliton crystals by the type of FWM involved opens the possibility of the generation of soliton crystals in single-pass systems. As they constitute soliton compounds with a strong regularity in their spatio-temporal periodicity, they relate to the generation of stable, low-noise frequency combs useful for applications in, for instance, spectroscopy or instrument calibration (cf. \cite{Zajnulina_2015}, \cite{Zajnulina_2017}). The existence of A-type breathers in optical fibers was experimentally shown in a series of papers by Vanderhaegen et al. \cite{Vanderhaegen_2020a}, \cite{Vanderhaegen_2020b}, \cite{Vanderhaegen_2021}. A similar or even the same setup configuration operated at higher input powers could be used to experimentally prove the existence of soliton crystals. 

    \textcolor{black}{The results of this study advance the fields of Nonlinear Optics and Wave Theory and are directly applicable to frequency-comb engineering and wave shaping, high-resolution spectroscopy, and optical systems, fibers (cf. \cite{Azana_2003}, \cite{Maram_2015}, \cite{Zhang_2015a}, \cite{Nikolic_2019}, \cite{Zajnulina_2024}) and semiconductors (cf.  Refs.~\cite{Wu_2022}, \cite{Wu_2023}), where controlling of nonlinear dynamics is critical.} 
    
    To illustrate the applicability of this study to an application, let us consider frequency-multiplexed optical computing. This study provides an answer to the question of what the primary driving mechanisms of data processing (FWM or solitons?) are in a frequency-multiplexed Extreme Learning Machine and explains the degradation of its performance at higher input powers as reported in Refs.~\cite{Zajnulina_2025}, \cite{Zajnulina_2025_Kerr}. When a frequency comb is used as an information carrier, data processing occurs through frequency-comb lines influencing each other, which happens via FWM \cite{Zajnulina_2025_Kerr} or, as shown here, XPM. As discussed in this study, the regime of separated solitons primarily relates to SPM, i.e., self-influencing, of frequency comb lines, while the information processing effects of FWM and XPM are negligible. As a result, the performance of a frequency-multiplexed Extreme Learning Machine deteriorates in the regime of separated solitons, which happens at higher input powers. Thus, FWM and XPM, rather than solitons, should be considered primary mechanisms in frequency-multiplexed optical computing. With it, an important question related to the system design and performance optimization of frequency-multiplexed optical computing schemes is solved. 

    Future work should include experimental validation of the regimes of the nonlinear temporal Talbot effect and their transitions. Additionally, studies of more realistic systems that incorporate effects such as higher-order dispersion, loss, and the Raman effect are necessary to better understand and exploit the regimes of the nonlinear temporal Talbot effect in real-world applications. \textcolor{black}{Whereas the current study provides an approximate analysis of the dynamics for low values of the phase modulation of the CW laser, $m\ll1$, future work should also consider higher values of $m$ and how they fit within the dispersion-related framework of the nonlinear temporal Talbot effect as presented here.}

\section*{Data availability} 
Data underlying the results presented in this paper are not publicly available at this time but may be obtained from me upon reasonable request.

\section*{Funding} \par 
Project Win4Space / Win4ReLaunch (SPW EER Wallonie Belgique, grant agreement number 2210181).

\section*{Disclosures} \par 
I declare no conflicts of interest.

\section*{Acknowledgments} \par 
I would like to say thank you to Michael B\"ohm (Wismar University of Applied Sciences, Germany), the inventor of Soliton Radiation Beat Analysis, for our long and inspiring conversations about optical solitons.




\end{document}